\documentclass[aps,pra,reprint,
twocolumn,longbibliography]{revtex4-1}
\usepackage{graphicx}
\usepackage{bm}
\usepackage{hyperref}
\usepackage{color}
\usepackage{dcolumn}
\usepackage{amsmath}
\usepackage{braket}
\usepackage{soul} % added by Philippe
\usepackage{xcolor}
\usepackage{epstopdf}

\begin{document}

%\preprint{APS/123-QED}

\title{Quantum and classical correlations in four-wave mixing from cold ensembles of two-level atoms}

\author{Lucas S. Marinho}
\affiliation{Departamento de F\'{i}sica, Universidade Federal do Piau\'{i}, Campus Ministro Petr\^{o}nio Portela, CEP 64049-550, Teresina, PI, Brazil.}

\author{Michelle O. Ara\'ujo}
\affiliation{Departamento de F\'isica, Universidade Federal de Pernambuco, 50670-901 Recife, Pernambuco, Brazil}

\author{Daniel Felinto}
\affiliation{Departamento de F\'isica, Universidade Federal de Pernambuco, 50670-901 Recife, Pernambuco, Brazil}

%\date{\today}

\begin{abstract}
Quantum correlations in four-wave-mixing from ensembles of cold two-level atoms may prevail without filtering over background light with well-known classical interpretations, such as Rayleigh scattering, as recently experimentally demonstrated in Phys. Rev. Lett. {\bf 128}, 083601 (2022). Here we provide an extended investigation of this effect, in which we detail the experimental procedure and the variation of the quantum correlation with various parameters of the system. Particularly, we show that the decay rate of the quantum correlations changes with the number of atoms in the sample, providing another indication of its superradiance-like nature. The nonclassical aspects of the signal occur for short timescales, but the long timescales carry as well a lot of information on the classical correlations of the system. This slow classical regime presents also two clearly distinct timescales, which we explain by two different pathways for the creation of biphotons. From the global analysis of the data in all its timescales, we are able to derive an empirical expression to fit the data, resulting in information on, among other parameters, the sample's temperature and superradiant-like acceleration. In general, the reported quantum correlations present a dependence on critical parameters of the system, such as optical depth and excitation power, that is quite different from other systems used for biphoton generation, and are more robust to changes in these parameters. This opens the possibility of exploring this process for efficient generation of narrowband biphotons or of other quantum-correlated photonic states of higher order.
\end{abstract}

%\pacs{Valid PACS appear here}

\maketitle

\section{Introduction}

An ensemble of two-level atoms coherently excited by optical fields is a basic model for radiation-matter interactions, being used as a first approximation for many physical systems close to a resonance \cite{Allen1987,Boyd2003}. Even though light scattered from single two-level atoms has shown quantum correlations since the early days of quantum optics~\cite{Kimble1977}, the observation of quantum correlations for light coming from ensembles of pure two-level atoms in free space has been elusive, with independent emissions from different atoms blurring the strong quantum signatures from individual atoms~\cite{Kimble1978, Loudon1983}. The problem is that the number of accidental coincidences in the correlation measurements grows with $N^2$, the square of the number of atoms, while coincidences originating from the same atoms grow with $N$ only.

\begin{figure}[h]
\centering
\includegraphics[width=8cm,height=5.0cm]{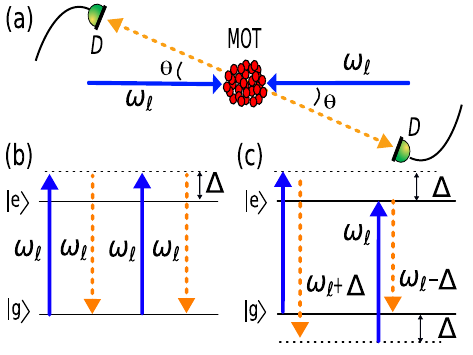}
\caption{(a) Scheme for spontaneous four-wave mixing (SFWM) from an atomic cloud. Two counterpropagating laser beams of frequency $\omega$ excite identical two-level atoms, and a counterpropagating pair of photons is generated forming an angle $\theta$ with the direction of excitation. Energy diagram of the two main processes participating in the generation of photon pairs: (b) Rayleigh scattering with the same frequency $\omega_{\ell}$ of the excitation fields and (c) scattering through sidebands located at detunes $\pm \Delta$ with respect to resonance, with $\Delta$ large when compared to the transition's natural linewidth.}
\label{fig1}
\end{figure}

A possible solution to this issue is to use a parametric nonlinear process to enhance the single-atom coincidences to a level above the accidental coincidences. The first nonlinear process explored to circumvent this problem was spontaneous four-wave mixing (SFWM)~\cite{Grangier1986}, in which two photons from excitation lasers are absorbed with each followed by the spontaneous emission of another photon in a way that, in the end, the system is left in the same initial state, see Fig.~\ref{fig1}(a). In this case, the conservation of energy and momentum is verified among the four photons involved in the process. The coincidence of detections of the two emitted photons results in a signal proportional to $N^2$ due to the constructive interference between the indistinguishable contributions of the four-photon process occurring in individual atoms in the ensemble. In this first experiment, the angle $\theta$ between excitation and detection was $90^{\circ}$, and an atomic beam was used as a source for a large number of atoms in the experiment's field of view. 

The cross-correlation function between the two photons in Ref.~\cite{Grangier1986} presented a symmetrical, oscillatory shape, with a minimum at zero delay. Such antibunching behavior was explained as a result of interference between sidebands, dislocated by the detuning to the atomic resonance [Fig.~\ref{fig1}(c)], and the background Rayleigh scattering, with the same frequency of the excitation laser [Fig.~\ref{fig1}(b)], that is the main cause of accidental coincidences. The creation of these sidebands was expected from previous studies on Mollow triplet in resonance fluorescence from single atoms~\cite{Mollow1969,Schuda1974,CohenTannoudji1979}. Recently, this Mollow triplet spectrum was observed from an ensemble of cold atoms \cite{OrtizGutierrez2019, Marinho2023}. The classical bound related to the autocorrelation function of one of the photon fields was also measured, but the cross-correlation between the fields of the two photons in the pair did not reach the threshold to demonstrate nonclassical correlations~\cite{Clauser1974}. Theoretical analysis was provided \cite{Grangier1986}, however, that showed the possibility of having cross-correlations of a purely quantum nature in the system.

Twenty years later, the problem of unfiltered SFWM in an ensemble of two-level atoms was revisited in a series of experiments with cold atoms~\cite{Kolchin2006,Du2007,Wen2007}. Again, no quantum correlations were observed, although the authors sought more systematically the violation of classical bounds in their signals and provided a more detailed account of the theoretical prediction for the observation of purely quantum correlations. 

Only in 2022 our group finally reported the experimental observation of such quantum correlations from unfiltered SFWM in a cold ensemble of pure two-level atoms~\cite{Araujo2022}. This observation was a consequence of the development of the research field on the observation and control of quantum correlations in four-wave mixing (FWM). Since the first observation of squeezed states in FWM in 1985~\cite{Slusher1985}, many groups have succeeded in observing various kinds of quantum correlations in FWM~\cite{Maeda1987,Raizen1987,Vallet1990,Lambrecht1996,Ries2003,McCormick2007}. Particularly important for the present work was the fast progress in generation and optimization of quantum correlations from SFWM after the proposal of the Duan-Lukin-Cirac-Zoller (DLCZ) protocol for long distance quantum communication~\cite{Duan2001}. The heart of the protocol is a delayed SFWM in two stages. First, a Raman transition, induced by a write pulse in an ensemble of three-level atoms in $\Lambda$ configuration, results in the storage of a collective entangled state in the system, heralded by the spontaneous emission of a single photon. Second, a read pulse maps the stored collective state into a second photon with high probability, using a resonant electromagnetically induced-transparency configuration. The overall SFWM process involved the absorption of two photons from the write and read fields and the spontaneous emission of the two corresponding photons. This proposal led to a fast experimental development, whose central aspect was the control of the stored collective states~\cite{Kuzmich2003,Balic2005,Matsukevich2005,Laurat2006,
Thompson2006,Zhao2009,Oliveira2014,Albrecht2015,
OrtizGutierrez2018}.

The relation between spontaneous emission and the creation of collective entangled states in atomic ensembles was famously established in the work by Dicke almost seventy years ago~\cite{Dicke1954}. Even though, since then the modeling of the role of spontaneous emission in the interaction of light with atomic ensembles has been largely dominated by semiclassical theories that neglect the creation of such collective states. Only after the DLCZ protocol, these entangled collective states start to gain increased attention, now associated with a class of phenomena without any alternative semiclassical explanation. A preliminary work for the observations of quantum correlations in Ref.~\cite{Araujo2022} was then the characterization of the classical bounds of correlation for the Rayleigh scattering from an ensemble of cold two-level atoms~\cite{Moreira2021}, in which it was theoretically shown that the first-principles theory for the problem involving collective states provides exactly the same result as the well known, and tested, semiclassical theory for thermal light sources composed of a large number of two-level atoms~\cite{Loudon1983}. In this previous work, it was also introduced experimental methods in order to deal with the typical nonergodic nature of the signals from cold atomic ensembles, coming from the limited periods of measurements between longer periods to replenish the atomic trap. 

Once one understands the role of collective states in the signals generated from two-level atoms, one can apply for them all techniques that were developed for $\Lambda$ systems in order to control such collective states. For example, it is well known that any spurious light or electromagnetic field kept on during measurements will disturb, to several degrees, the phase between atoms in the collective state, smoothing signals relying on the state's coherent nature~\cite{Felinto2005}. The expected violation of classical bounds by unfiltered two-level systems is quite small, on the order of fractions of the bound value~\cite{Grangier1986,Du2007,Wen2007}, making them extremely sensitive to the smallest disturbances by external fields or misalignments in the experimental setup. In contrast, filtered $\Lambda$ systems presented violations larger than hundreds of the value of the classical bounds~\cite{Laurat2006}. Even though, the introduction of filters in the signals from two-level systems should enhance their nonclassical aspects~\cite{Aspect1980}, opening the use of strong cycling transitions for diverse applications in quantum information, from quantum sensing to the efficient generation of quantum correlated fields for quantum communication. An initial step was taken by us in this direction, where we achieved a proof of the principle of increasing correlations from spectral filtering \cite{Marinho2023}. 

The present work extends the one in Ref.~\cite{Araujo2022} in various aspects. First, in Sec.~\ref{sec_setup}, we provide a deeper account of its methodology. In Sec. \ref{sec:setup_timing} we give some details about our experimental setup followed by the data analysis procedure (Sec. \ref{sec:data_analysis}). The main source of the nonergodicity nature of our signal is investigated in Sec. \ref{sec:non_ergodicity}. The normalized correlation functions and Cauchy-Schwarz inequality are introduced in Sec. \ref{sec:normalized_correlation} to quantify the quantum behavior of the system. A discussion, with supporting experimental results, on how the quantum correlation degrades with the relaxation of some of the experimental conditions is included in Appendix~\ref{ap1}. Second, in Sec.~\ref{sec_fast}, the empirical model presented in Ref.~\cite{Araujo2022} is revisited (Sec. \ref{sec:fast_empirical_expression}) and we provide new measurements of the correlation decay as a function of the number of atoms in the sample, which reinforces the indication of its superradiant nature coming from the formation of collective states in the medium (Sec. \ref{sec:superradiance}). We also introduce in Sec. \ref{sec:g2c} the first measurements of a third-order correlation function in the system, with beatings in the conditioned autocorrelation functions being considerably reinforced with respect to the ones observed in the unconditioned signals. Third, in Sec.~\ref{sec_slow},  we introduce a new experimental setup, with a single beam exciting the sample, that helps to clarify the origin of long-time correlations in the system (Sec. \ref{sec:single_beam}). We also provide a series of measurements for the long-term correlations in the system. Even though these slow correlations are all within the classical bounds of our system, they provide considerable information on its dynamics, such as direct measurements of the ensemble's temperature and the role of different scattering angles on the observed correlations. In Sec. \ref{sec:complete_empirical_expression}, we extend our empirical model to all timescales.  Finally, in Sec.~\ref{sec_conclusion} we draw our conclusions.

\section{Experimental procedure}
\label{sec_setup}

In this section, we recall our experimental setup \cite{Araujo2022} and provide a detailed discussion about the measurement procedure, including the data analysis and ensemble averages.

\subsection{Setup and timing}\label{sec:setup_timing}

\begin{figure}[h]
\centering
\includegraphics[scale=1]{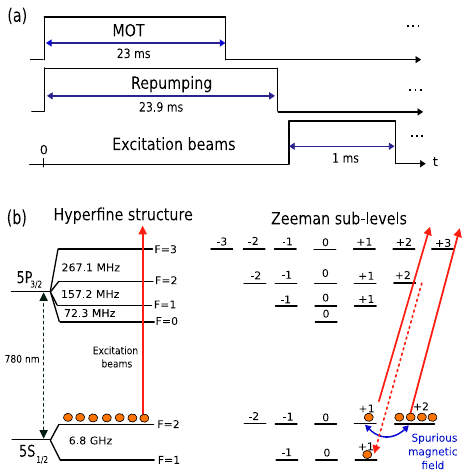}
\caption{(a) Time sequence of the experiment. (b) Hyperfine and Zeeman structure of the levels participating in the FWM process and the spurious optical pumping mechanism to the dark $5S_{1/2}(F=1)$ level.}
\label{fig2}
\end{figure}

An atomic cold cloud of $^{87}$Rb atoms is prepared by turning on the trapping lights, repumping lights, and magnetic fields in a magneto-optical trap (MOT) during $23$ ms [Fig. \ref{fig2}(a)]. The trapping beams and magnetic field are then turned off, but the repumping laser is kept on during an additional $0.9$ ms to optically pump the atoms to the $5S_{1/2}(F=2)$ hyperfine ground state [Fig. \ref{fig2}(b), left]. After $23.9$ ms, all lasers and magnetic field are off and we illuminate the cloud for 1 ms with excitation laser beams blue detuned by $\Delta$ with respect to the transition from $5S_{1/2}(F=2)$ to $5P_{3/2}(F^{\prime}=3)$, with a wavelength around 780 nm. These excitation beams are circularly polarised and optically pump the atoms inside the $5S_{1/2}(F=2)$ manifold to the extreme Zeeman state $5S_{1/2}(F=2,m_F=+2)$~\cite{Araujo2022SM}, from which the atoms are constrained to the cycling transition $5S_{1/2}(F=2,m_F=+2) \rightarrow 5P_{3/2}(F^{\prime}=3,m_{F^{\prime}}=+3)$ [Fig. \ref{fig2}(b), right]. In this situation, the atoms in the ensemble can be well approximated as pure two-level systems with a decay rate of $\Gamma = 2\pi\times 6.06$ MHz~\cite{Steck2001}.

The typical optical depths $OD$ in our cold ensemble range from 4 to 18, and are controlled by tuning the power of the trapping beams~\cite{Oliveira2014,Moreira2021} and measured by determining the detuning of the probe light resulting in the transmission of half the initial pulse~\cite{Gattobigio2010}. We estimate that $OD = 15$ implies about $N = 10^6$ atoms in the region of the ensemble corresponding to the detected mode~\cite{Oliveira2014}. The temperature of the cold atomic cloud is around hundreds of $\mu$K~\cite{Moreira2021}.

\begin{figure}[h]
\centering
\includegraphics[scale=1]{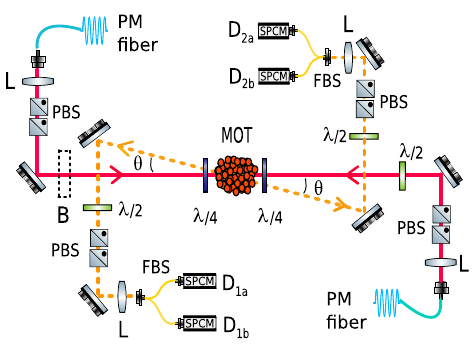}
\caption{Experimental setup for spontaneous four-wave mixing. The counterpropagating excitation beams are represented by the solid lines in red. Detected modes are indicated by the dashed orange lines making an angle $\theta$ with the excitation direction. The dashed rectangle $B$ is placed in the position where one of the excitation fields is blocked for the experiments involving excitation coming from a single direction. PM fiber, polarisation maintaining fiber; MOT, magneto optical trap; PBS, polarizing beam splitter; L, lens; FBS, fiber beam splitter; SPCM, single photon counting module; $\lambda$/2 ($\lambda$/4), half wave plate (quarter wave plate); $D_i$, detector $i$, with $i = $ 1a, 1b, 2a, or 2b.}
\label{fig3}
\end{figure}

Most of the experiments we report here were performed with two counterpropagating excitation beams, but in some experiments, one of the excitation beams was blocked (Fig.~\ref{fig3}). The beams were generated using polarization maintaining (PM) optical fibers and focused to a $4\sigma$ diameter of $420$ $\mu$m on the atomic ensemble. They have the same power $P$, controlled by independent polarizing beam splitters (PBS) and half wave plates ($\lambda/2$), and the same frequency, as they are derived from the splitting of the same laser beam. The circular polarization is set by quarter wave plates ($\lambda/4$) right before the windows of the vacuum chamber. Before reaching the quarter wave plates, the linear polarizations of the excitation beams are cleaned by sequences of two polarizing beam splitters (PBS) in each arm and one of the beams is rotated by a half wave plate to ensure they have the same circular polarization in the atomic reference frame of the ensemble. 

The two emitted photons are collected in a direction making an angle of $\theta=3.0 \pm 0.3^{\circ}$ with respect to the excitation beams and are directed to independent single-mode fibers  (Fig.~\ref{fig3}). The optical mode coupled in the detection fibers has a diameter of  $140$ $\mu$m on the atomic ensemble. The detected mode is aligned to pass in the middle of the excitation beams, as seen on the two CCD cameras used to monitor the cloud of cold-trapped atoms. The two optical fibers for the excitation beams are coupled to each other with efficiencies larger than $70\%$, with similar coupling efficiency observed between the two detection fibers. The quarter wave plates on the two sides of the vacuum chamber turn the circular polarization of the emitted photons into linear polarization, with a typical 99$\%$ degree of polarization checked right before the detection inputs. This degree of polarization for the emitted photons provides our simplest verification for the polarization of the atoms in the intended two-level transition and see Supplemental Material of Ref.~\cite{Araujo2022} %in Ref.~\cite{Araujo2022SM} 
we show a more rigorous technique through microwave spectroscopy to characterize the populations of the atoms inside the $5S_{1/2}(F=2)$ manifold. The polarizations of the excitation beams and detected modes are all optimized to maximize this number.

Four detectors ($D_{1a}$,$D_{1b}$ for field 1 and $D_{2a}$,$D_{2b}$ for field 2) collect the emitted photons during the $1$-ms excitation window (Fig.~\ref{fig3}), with each detected mode split in two by a fiber beam splitter (FBS) right before the respective detectors. In this way, we have a Hanbury-Brown-Twiss interferometer in each detected mode, allowing us to obtain the second-order autocorrelation function for each field~\cite{Loudon1983} simultaneously to the cross-correlation functions between the two fields. The detectors are avalanche photo-detectors (APD, model SPCM-AQRH-13-FC from Perkin Elmer). In order to evaluate these correlation functions, the detection events are recorded using a multiple-event time digitizer (model MCS6A from FAST ComTec).

\subsection{Data analysis}\label{sec:data_analysis}

The time digitizer generates a file containing all the instants $t$ in which each APD fired, where $0 \leq t \leq 1$ ms with a resolution of $100$ ps. We remove then the data acquired during the first $1$ $\mu$s and the last $20$ $\mu$s, because they lay at the rising and falling edges of the $1$ ms pulses. We also remove any possible afterpulse by neglecting other detections occurring in a time window of 100 ns after the firing of that detector~\cite{Cova1996}. 

After these preliminary actions, our analysis program scans the files in order to collect single and joint counts. We compute then the quantities $N_i(t)$, $N_{ij}(t,t+\tau)$, and $N_{\text{tot}}$, where

%\begin{enumerate}[(i)]

%\item
(ii) [$N_i(t)$] are the single counts of the detector $i$ at instant $t$, i.e., the number of times that the detector $i$ fired at $t$. Numerically, the $N_i(t)$ are vectors containing $n$ rows ranging from $0 \leq t \leq 1$ ms in steps of $0.1$ ns.

%\item
(ii) [$N_{ij}(t,t+\tau)$] is the joint coincidence between the detectors $i$ and $j$, i.e., the number of times that detector $j$ fired at $t+\tau$ after the detector $i$ fired at $t$. Numerically, the $N_{ij}(t,t+\tau)$ are matrices with $n$ rows and $m$ columns ranging from, respectively $0\leq t \leq 1$ ms with steps of $0.1$ ns and $0 \leq \tau \leq \tau_{max}$ also with steps of $0.1$ ns, where $\tau_\mathrm{max}$ is set previously.

%\item
(iii) [$N_\mathrm{tot}$] is the total number of samples, i.e., the number of times that the $1$ ms-pulses were sent to the cloud. As each cycle has a duration of $25$ ms [cf. Fig. \ref{fig2}(a)], for a total acquisition time of $30$ minutes for a single file, we have a total of $72\,000$ cycles per file. For several files, we sum the $N_i$, $N_{ij}$, and $N_\mathrm{tot}$ evaluated separately.

%\end{description}

From the $N_i(t)$, $N_{ij}(t+\tau)$, and $N_
\mathrm{tot}$ above, we compute the probabilities of single and joint counts $p_i(t)$ and $p_{ij}(t,t+\tau)$, respectively, as

\begin{equation}
p_i(t)=\dfrac{N_i(t)}{N_\mathrm{tot}} \label{eq_pi}
\end{equation}
and
\begin{equation}
p_{ij}(t,t+\tau)=\dfrac{N_{ij}(t,t+\tau)}{N_\mathrm{tot}}. \label{eq_pij}
\end{equation}

\subsection{Optical pumping and non-ergodicity}\label{sec:non_ergodicity}

Once we have initially prepared the atoms in the $5S_{1/2}(F=2,m_F=+2)$, unfortunately, they do not remain indefinitely in the cycling transition, and we do observe a slow optical pumping in the system to the dark $5S_{1/2}(F=1)$ level. We understand this optical pumping as coming mainly from residual magnetic fields that transfer some atoms to the $5S_{1/2}(F=2,m_F=+1)$, from which they have a small probability of being excited to $5P_{3/2}(F^{\prime}=2,m_{F^{\prime}}=+2)$ and spontaneously decay to $5S_{1/2}(F=1)$, a process illustrated on the right of Fig.~\ref{fig2}(b). In order to minimize this residual magnetic field, we use the microwave spectroscopy technique described in Refs.~\cite{Araujo2022, Moreira2021, Almeida2016}. In this way, we cancel these residual magnetic fields down to around 23$\,$mG. Even so, the effect is only reduced but not totally eliminated.

\begin{figure}[h]%[!ht]
\centering
\includegraphics[width=8.5 cm,height= 9.0 cm]{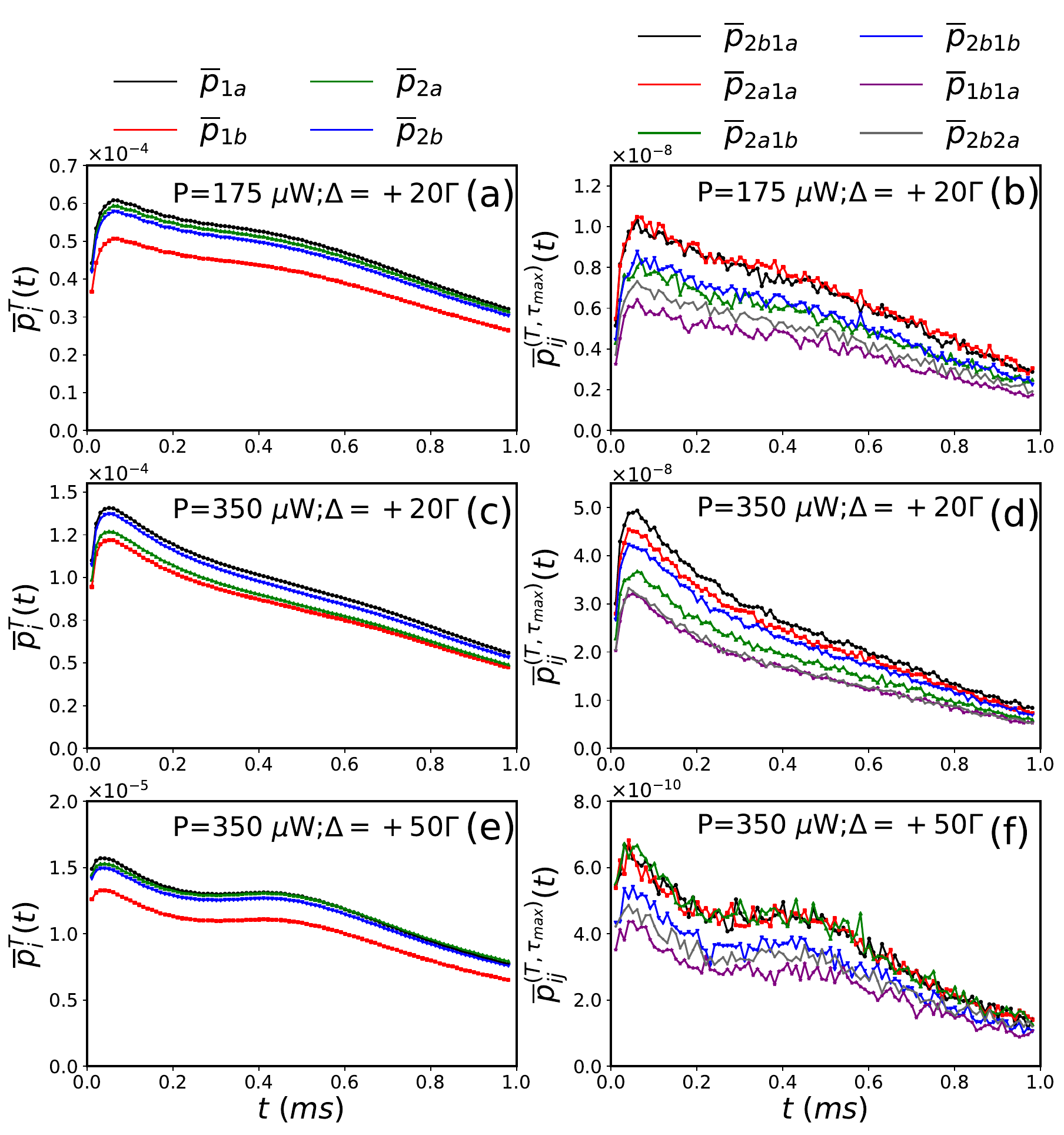}
\caption{Probabilities as a function of time for different optical excitation powers $P$ and detunings $\Delta$. On the left-hand side, plots (a), (c), and (e), probabilities of singles $\bar{p}_i^{T}(t)$. On the right-hand side, plots (b), (d), and (f), the corresponding maximum join probabilities $\bar{p}_{ij}^{(T,\tau_{max})}$, with averages over a time interval of $T=10$ $\mu$s around $t$. }
\label{fig4}
\end{figure}

To make the data analysis consistent, it is crucial to take into account the typical nonergodic nature of our signal, i.e., processes where the atomic sample is not stationary over time, either during the $1$-ms period \cite{Araujo2022} or due to the long-term MOT fluctuations during the whole acquisition time \cite{Moreira2021}. The optical pumping described above results in a reduction in the number of atoms in the ensemble during the excitation period, as shown in Fig. \ref{fig4}, through the decay of $p_i$ and $p_{ij}$ over time. The MOT fluctuations are due to environmental random factors such as room temperature and atom number fluctuations due to long-time fluctuations of our atomic source.

Figure \ref{fig4} exhibits the behavior of probabilities as a function of time for different optical excitation powers $P$ and detunings $\Delta$. To obtain these probabilities, first, we compute them using Eqs. (\ref{eq_pi}) and (\ref{eq_pij}) for each time $t$ over all
trial periods and later we average them over a time interval of $T=10$ $\mu$s around $t$.
For the plots on the right-hand side, in each of them, we plotted their maxima
with respect to variations of $\tau$ for a particular instant $t$, giving the quantity $\bar{p}_{ij}^{(T,\tau_{max})}$. Note that the system does not reach a steady state, with the probabilities falling down over time. On the other hand, the single probabilities are on the order of $10^{-4}$ while the joint probabilities depend on the square of this value, i.e., $10^{-8}$. Then, a long period of measurements is required in order to obtain a high signal-to-noise ratio, causing our signal to be susceptible to the long-term fluctuations. These effects are treated appropriately through the ensemble averages  (\ref{eq_pi}) and (\ref{eq_pij}).

\subsection{Normalized correlation functions}\label{sec:normalized_correlation}

From the single and joint probabilities of Eqs. (\ref{eq_pi}) and (\ref{eq_pij}) above, we finally compute the normalized second-order correlation functions

\begin{equation}
\label{eq_gij_matrix}
g_{ij}(t,t+\tau)=\dfrac{p_{ij}(t,t+\tau)}{p_i(t)p_j(t+\tau)},
\end{equation}
where $g_{ij}$ is the correlation between detectors $i$ and $j$. This method of analysis assures that all $g_{ij}(\tau)\rightarrow 1$ for $\tau\rightarrow\infty$ (uncorrelated limit). Numerically, the product $p_i(t)p_j(t+\tau)$ is a matrix of the same size as $p_{ij}(t,t+\tau)$, and the division in Eq.(\ref{eq_gij_matrix}) is taken element by element. As we have four detectors, and considering all possibilities of click arrangements, we have six correlation functions $g_{ij}=\lbrace g_{1a2a}, g_{1a2b}, g_{1b2a}, g_{1b2b}, g_{1a1b}, g_{2a2b} \rbrace$, and six correlation functions $g_{ji}=\lbrace g_{2a1a}, g_{2b1a}, g_{2a1b}, g_{2b1b}, g_{1b1a}, g_{2b2a} \rbrace$. Since $\bar{g}_{ij}$ and $\bar{g}_{ji}$ differ only by the order of the detector which fired first, we can build a single correlation function, denoted by $g_{ij}$ for $\tau<0$ and $\tau>0$, where $g_{ij}$ ($g_{ji}$) is valid for $\tau<0$ ($\tau>0$). Thus, we have a total of six correlation functions from $-\tau$ to $\tau$. Four of these $g_{ij}$ functions are called cross-correlation functions because they account for the correlations between photons at different fields ($1$ and $2$), and we denote them by $g_{12}$. Two of the $g_{ii}$ functions are called autocorrelation functions because they account for the correlations between photons of the same field ($1$ or $2$), and we denote them by $g_{11}$ and $g_{22}$. The $g_{11}$ and $g_{22}$ have features of thermal fields~\cite{Eloy2018,Moreira2021} with $g_{11}(0)=g_{22}(0)=2$. 

Our criterion for searching quantum correlations is the violation of a Cauchy-Schwarz inequality~\cite{Clauser1974} valid for classical fields:

\begin{equation}
\label{eq_R}
R(\tau)=\dfrac{g_{12}(\tau)^2}{g_{11}(0)g_{22}(0)} \leq 1,
\end{equation}
in the ideal situation where all detectors have the same efficiency. In practice \cite{Clauser1974}, for our setup, we have two $R$ parameters, $R_1$ and $R_2$, given by
\begin{gather}
    R_1(\tau) = \dfrac{g_{1a2b}(\tau) g_{1b2a}(\tau)}{g_{1a1b}(0) g_{2a2b}(0)}, \; \;
     R_2(\tau) = \dfrac{g_{1a2a}(\tau) g_{1b2b}(\tau)}{g_{1a1b}(0) g_{2a2b}(0)}.  \label{R1_R2}
\end{gather}

\begin{figure}[h]
\centering
\includegraphics[width=8.5 cm,height= 9.0 cm]{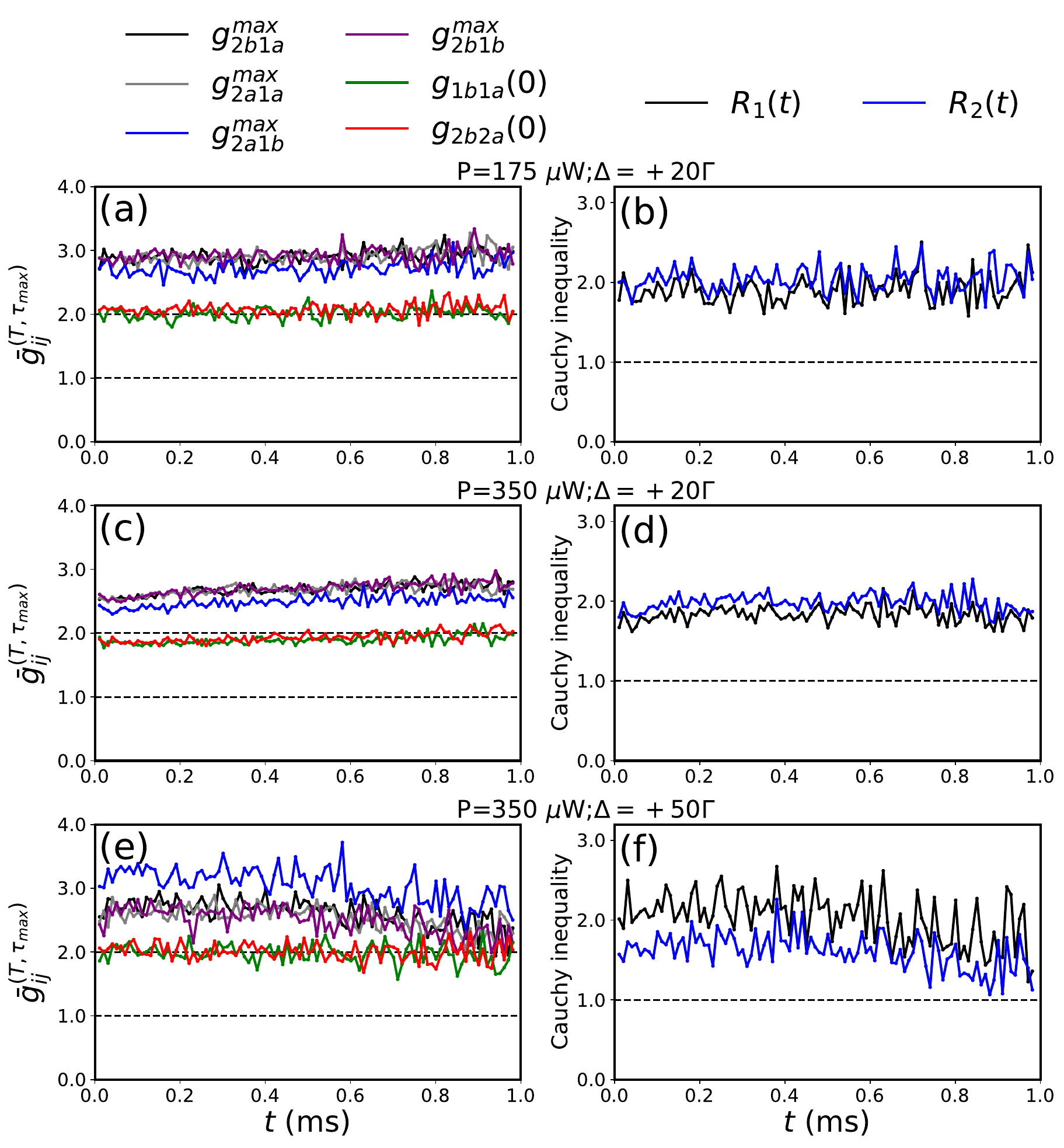}
\caption{Maxima of the normalized correlation functions [(a), (c), and (e)], and Cauchy-Schwarz inequalities [(b), (d), and (f)] as a function of time for different optical excitation powers $P$ and detunings $\Delta$, with averages over a time interval of $T=10$ $\mu$s around $t$.}
\label{fig5}
\end{figure}

Figure \ref{fig5} show the maxima of the normalized correlations functions $\bar{g}_{ij}^{(T,\tau_{max})}$  and Cauchy-Schwarz parameter $R_1$ and $R_2$, as a function of time. For each point, we take averages over a time interval of $T=10$ $\mu$s around $t$. The correlation functions were obtained directly from the probabilities in Fig. \ref{fig4} and applying the definition (\ref{eq_gij_matrix}). We see that although the probabilities decay with time, the correlations and the Cauchy-Schwarz inequalities remain nearly constant throughout the measurement period. Dashed lines, on the left-hand side, indicate levels 1 for uncorrelated fields and 2 for thermal fields. Note that, as expected, $g_{11}(0)= g_{22}(0)\approx 2$ \cite{Moreira2021}. On the right-hand side, the dashed lines indicate the classical bound of the Cauchy-Schwarz criterion. We see a clear inequality violation during the whole period, showing its nonclassical behavior and robustness, even varying the optical power and detuning of the excitation beams.

\begin{figure}[h]
\centering
\includegraphics[width=8.5 cm,height= 8.5 cm]{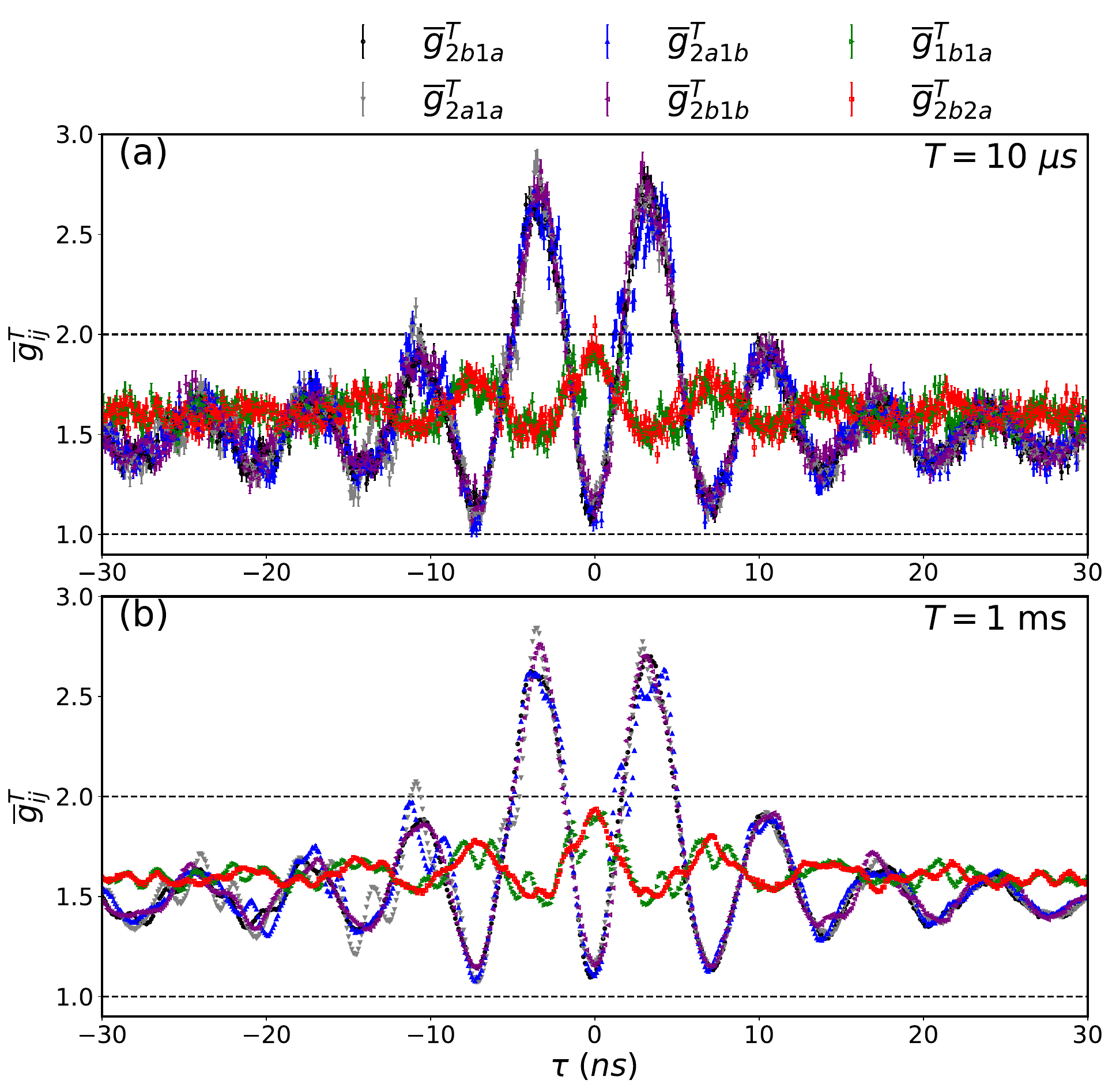}
\caption{Correlations functions $\bar{g}_{ij}^{T}(\tau)$ as a function of $\tau$ considering an average over (a) $T=10$ $\mu$s around the instant $t=0.5$ ms and (b) $T=1.0$ ms. Other parameters were excitation power $P=350$ $\mu$W  and detuning $\Delta = +20\Gamma$. }
\label{fig6}
\end{figure}

The behavior of the correlations functions $\bar{g}_{ij}^{T}(\tau)$ as a function of $\tau$ is shown in Fig. \ref{fig6}. The presence of oscillations and antibunching in these correlation functions were reported and discussed previously \cite{Du2007,Wen2007,Araujo2022}, and are due to the interference between the two emission processes in the SFWM (interference between processes (b) and (c) in Fig. \ref{fig1}). In Fig. \ref{fig6}(a), we considered a time average only over  $T=10$ $\mu$s around the instant $t=0.5$ ms, while in (b) the average was taken over the whole period ($T=1.0$ ms). We highlight here the consistency of using the time average approach, which improves statistics while preserving all features of the signal. Therefore, from now on we will always consider an average with $T=1.0$ ms and omit this symbol.  As discussed in Ref.~\cite{Moreira2021}, this procedure of computing the correlation functions allows us to treat nonergodic processes and enhance our data statistics.

\section{Fast quantum regime}
\label{sec_fast}
\begin{figure}[h]
\centering
\includegraphics[width=8.5 cm,height= 8.5 cm]{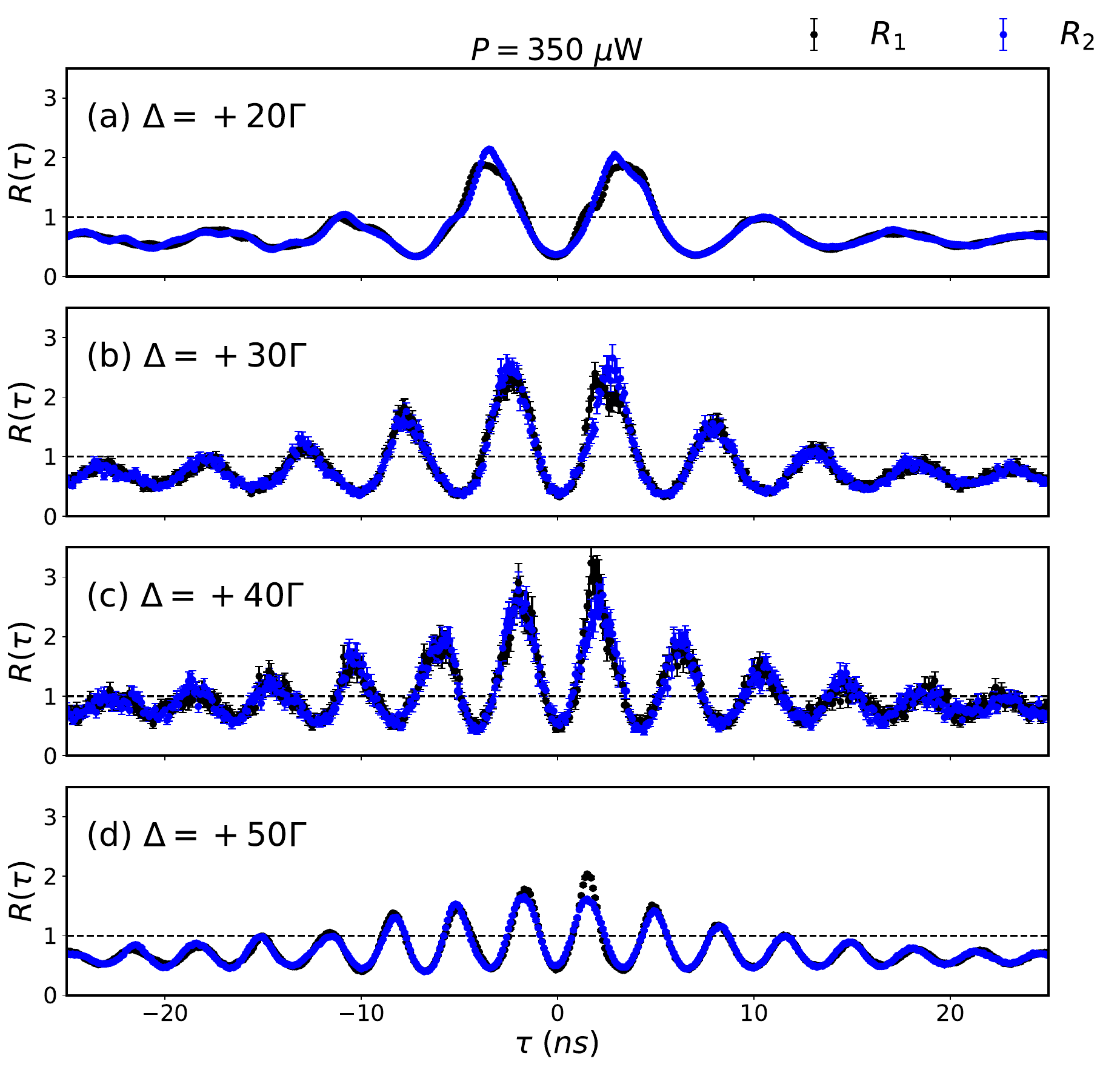}
\caption{Dependence of the parameters $R_1$ and  $R_2$ of the Cauchy-Schwarz criterion as a function of $\tau$. The power was kept constant at $P = 350$ $\mu$W while we varied the detuning, with (a) $\Delta = +20\Gamma$, (b) $\Delta = +30\Gamma$, (c) $\Delta = +40\Gamma$, and (d)$\Delta = +50\Gamma$. The dashed line indicates the frontier between quantum and classical correlations.}
\label{fig7}
\end{figure}

In this section, we focus on the correlation analysis
in the fast quantum regime of tens of nanoseconds, the
region where the Cauchy-Schwarz criterion is violated. Figure \ref{fig7} exhibits the time behavior of the $R_1$ and $R_2$ parameters varying the detuning while keeping constant the excitation power. We observe that, for all detunings, the oscillation persists approximately until 25 ns, but the maximum amplitude and the violation region vary for each detuning. Furthermore, the delay $\tau_{max}$ that corresponds to the maximum value of the inequality violation is inversely proportional to the detuning $\Delta$ of the atomic transition, i.e., $\tau_{max} \propto 1/\Delta$. 

In addition to the dependence of the correlations with the repumping beam and the polarization of the excitation beams (see Appendix \ref{ap1}), the criterion violation value is greatly affected by the alignment of the four-wave mixing, which depends on the interaction region of the excitation beams and the detection modes being exactly in the center of the cloud of cold atoms. Also, the fiber couplings of the excitation and detection modes must be on the order of 70 to 80\%.

\subsection{Fast empirical expression}\label{sec:fast_empirical_expression}

At this point, we will analyze in greater detail how the theoretical model \cite{Du2007,Wen2007} can be modified empirically to provide us with a satisfactory description of the temporal behavior of the correlation functions. In the limit of large detunings, short delays, and low
excitation power, the correlations can be written as \cite{Du2007}

\begin{equation}\label{gij_Wen}
    g_{12}(\tau) = 1+ \frac{4}{\pi^2}\Big[ e^{-2\gamma_g\tau} + e^{-\gamma_e\tau} - 2e^{-(\gamma_g  +\gamma_e)\tau}\cos(\Delta\tau) \Big],
\end{equation}
where $\gamma_e$ ($\gamma_g$) is the decay rate of the excited (ground) state and $\Delta$ is the detuning from the atomic resonance. To match the experimental data with theory, we assumed that $\gamma_g = 0$ and empirically modified Eq. (\ref{gij_Wen}) to \cite{Araujo2022}

\begin{equation}\label{gij_empirical}
   g_{12}(\tau) = 1+ f\frac{4}{\pi^2}\Big[ 1 + e^{-\chi\Gamma\tau} - 2e^{-\chi\Gamma\tau/2}\cos(\Delta_{fit}\tau) \Big],
\end{equation}
with $f$, $\chi$, and $\Delta_{fit}$ being fit parameters.

\begin{figure}[h]
\centering
\includegraphics[width=8.5 cm,height= 5 cm]{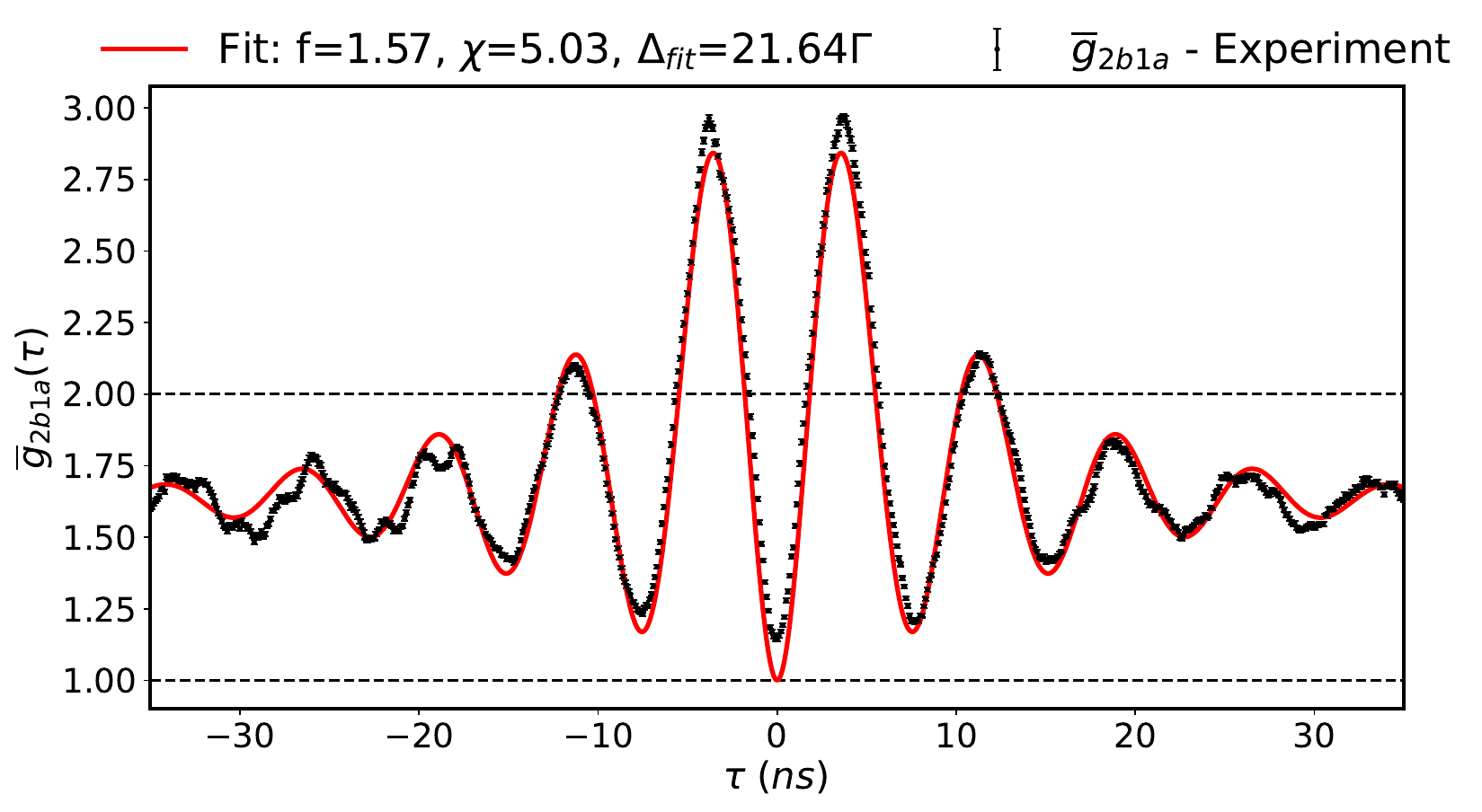}
\caption{Correlation function $\bar{g}_{2b1a}^{T}(\tau)$ (filled black circles) and theoretical empirical fit (solid red line) from Eq. (\ref{gij_empirical}). Measurement was carried out with $P = 175$ $\mu$W and $\Delta = +20\Gamma$. Similar results were obtained for all other cross-correlation functions.}
\label{fig8}
\end{figure}

The red solid line in Fig. \ref{fig8} provides a fit from Eq. (\ref{gij_empirical}) for a measurement with $P = 175$ $\mu$W and $\Delta = +20\Gamma$. As a result, we obtain $f = 1.57 \pm 0.01$, $\chi = 5.03 \pm 0.05$, and $\Delta_{fit} = (21.64 \pm 0.02)\Gamma$. The value of $\Delta_{fit} \approx \Delta$ is consistent with the theoretical expectation. The value $f \neq 1$, on the other hand, indicates that the simplified theory of Refs. \cite{Du2007,Wen2007} does not accurately predict the maximum value of the correlation functions. It is worth highlighting that the maximum theoretical values for the correlation functions obtained in these references do not coincide with each other, with Ref. \cite{Du2007} giving $f = \pi^2/4$ and Ref. \cite{Wen2007} giving $f = 1$. Another striking difference was an increase in the rate of natural decay, represented by the value $\chi \geq 1$, where for an individual atom $\chi = 1$. Therefore, this may indicate the existence of an acceleration in the decay rate due to collective effects, as observed in Refs.~\cite{OrtizGutierrez2018,Araujo2016} under similar experimental conditions.

\subsection{Superradiance-like effects}\label{sec:superradiance}

\begin{figure}[h]
\centering
\includegraphics[width=8.5 cm,height= 8.5 cm]{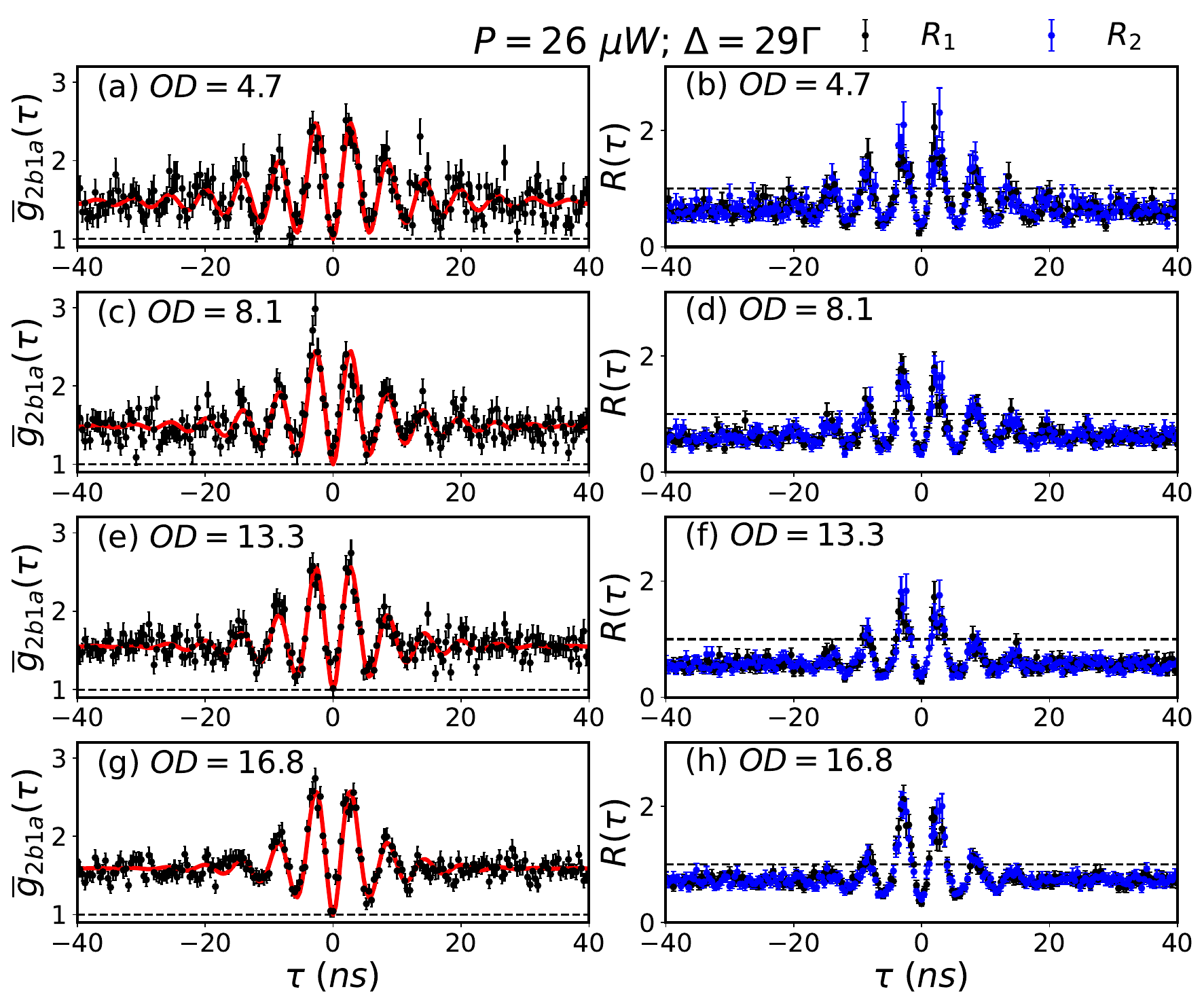}
\caption{On the left-hand side, we show the correlation function $\bar{g}_{2b1a}^{T}(\tau)$ (filled black circles) and theoretical empirical fit (solid red line) from Eq. (\ref{gij_empirical}) for different Optical Depths $OD$. On the right-hand side, are the corresponding Cauchy-Schwarz inequalities. }
\label{fig9}
\end{figure}

In order to study the fast acceleration decay rate observed in our data, we exhibit in Fig. \ref{fig9} the correlation functions and Cauchy-Schwarz inequalities for various optical depths, keeping excitation power and detuning fixed at $P = 25$ $\mu$W and $\Delta = +29\Gamma$, respectively. The solid red lines are theoretical empirical fits from Eq. (\ref{gij_empirical}). As the optical depth increases, we can see an increase in the decay rate, characterized by the oscillations ``dying" in progressively shorter periods of time as $OD$ increases. Also, note that the maximum values of the Cauchy-Schwarz criterion are practically independent of variations in the number of atoms, which is in line with the theoretical prediction \cite{Du2007,Wen2007}. As the correlation function can be written as $g_{12}(\tau)=1+R_{cc}(\tau)/R_R^2$, where the coincidence rate $R_{cc} \propto N^2$ (as expected for parametric processes) and $R_R \propto N$ represents the Rayleigh scattering rate, this type of behavior is quite different from what typically occurs in type $\Lambda$-systems, where correlations grow with the increase in the number of atoms \cite{Duan2001,Laurat2006,Oliveira2014}. 

\begin{figure}[h]
\centering
\includegraphics[width=8.5 cm,height= 5 cm]{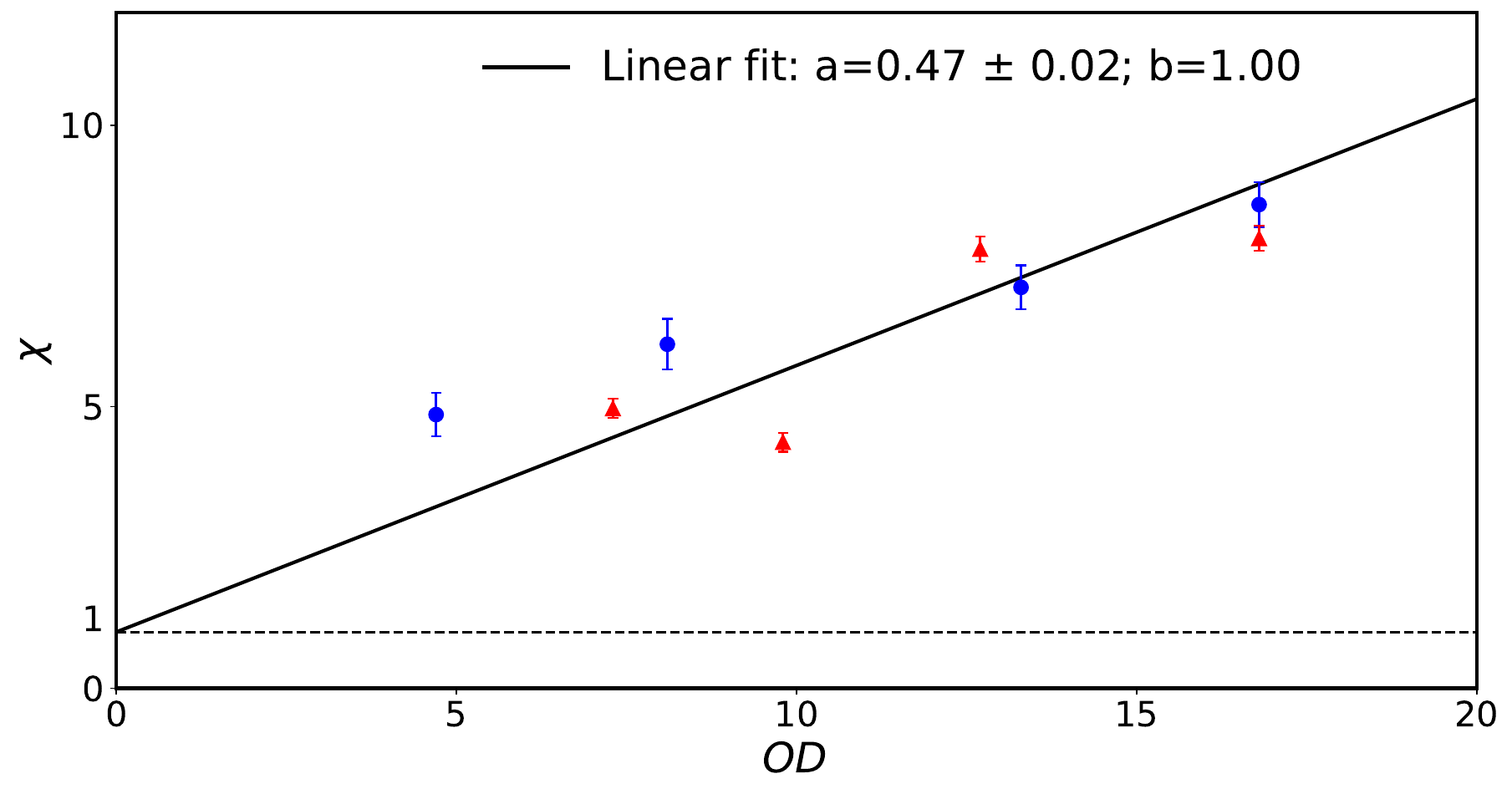}
\caption{Dependence of the $\chi$ parameter with the optical depth. Measurements with fixed power $P = 25$ $\mu$W but detunings $\Delta=+29\Gamma$ (filled blue circles) and  $\Delta=+20\Gamma$ (filled red triangles). The solid black line is a global linear fit, $\chi = a \times OD + b$, of the data.  }
\label{fig10}
\end{figure}

The dependence of the $\chi$ parameter on the optical depth is plotted in Fig. \ref{fig10}. The filled blue circles correspond to measurements with detuning $\Delta=+29\Gamma$, and the filled red triangles correspond to measurements with the same power of $P = 25$ $\mu$W but with detuning $\Delta=+20\Gamma$. It is important to remind here that the number of atoms is not fixed throughout our 1-ms excitation period, as can be inferred from the decays in Figs. \ref{fig4}(a), \ref{fig4}(c), and \ref{fig4}(e). The fitted $\chi$ reflects then an average acceleration of the decay rate in that interval. We observe a linear relation of $\chi$ with $OD$, in addition to its independence with the detuning $\Delta$. A global linear fit, $\chi = a \times OD + b$, of the data resulted in $a = 0.47 \pm 0.02$, where we fixed $b=1.00$ to ensure that there is no acceleration in the decay rate in the limit of a very thin sample. This behavior in decay rates is consistent with superradiance-like effects \cite{Oliveira2014,Araujo2016,Roof2016, OrtizGutierrez2018}. Such accelerated decay can appear from the formation of a symmetric collective state in the excited state
of the atoms in the ensemble. The mechanism is likely related to the time ordering for the formation of sidebands in Fig. \ref{fig1}(c), which was recently directly observed in Refs. \cite{Marinho2023, Masters2023}. As pointed out a long time ago in Ref. \cite{CohenTannoudji1979} (see Fig. 7 of this reference), the first photon to leave the medium in the SFWM in an ensemble of two-level atoms is the one out of resonance. The ensemble is then left with a single atom in the excited state, which will later decay into the second photon. In this case, the first photon heralds the occurrence of an instantaneous three-photon process that leaves an atom
in the excited state \cite{Masters2023}. With the large detunings employed in our experiments, the absorption of the excitation fields is negligible throughout the ensemble, resulting in a uniform, equal probability to excite any atom in the ensemble. We expect this mechanism to result in a symmetrical collective state in the excited state, such as the one resulting in the superradiant emission in the reading process of a DLCZ quantum memory \cite{Oliveira2014}.

\subsection{Conditioned second-order correlations}\label{sec:g2c}

\begin{figure}[h]
\centering
\includegraphics[width=8.5cm,height=8.5cm]{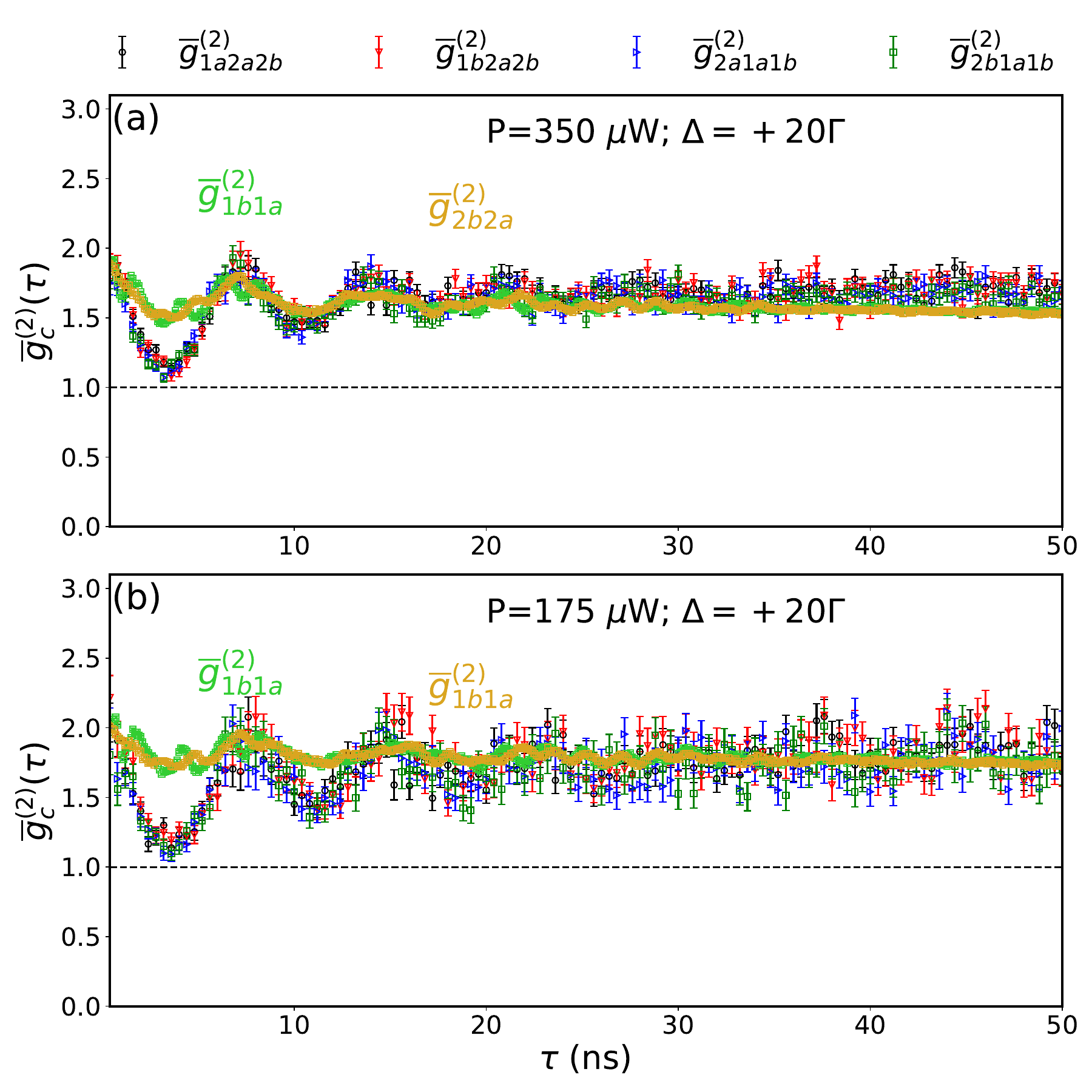}
\caption{Conditioned second-order correlation function $g_{c}^{(2)}(\tau)$ for $\Delta=20\Gamma$ and excitation powers (a) $P=350$ $\mu$W and (b) $P=175$ $\mu$W. The dashed black line denotes the limit of inequality $g_{c}^{(2)} \geq 1$, valid for classical fields. For comparison, we added the second-order correlations functions, $g^{(2)}_{1b1a}$ (lime green) and  $g^{(2)}_{2b2a}$ (gold).}
\label{fig11}
\end{figure}

The Cauchy-Schwarz inequality is not the only criterion for searching for quantum correlations in optical systems. Another well-known condition used to characterize the single-photon regime is the so-called conditioned second-order correlation $g_{c}^{(2)}(\tau)$ function~\cite{Loudon1983,Oliveira2014}, denoted by $g_{122}=g_{ijj'}$ and defined as

\begin{equation}
g_{ijj'}(t,t+\tau)=\dfrac{p_{ijj'}(t,t+\tau)}{p_{ij}(t)p_{ij'}(t+\tau)} p_i(t),
\label{eq_g2c}
\end{equation}
where $p_{ijj'}(t,t+\tau)$ is the probability of triple coincidences, i.e., the probability for the detectors $j$ and $j'$ of the same field to click together at the instant $t+\tau$ after one of the detectors of the opposite field have clicked at the instant $t$ [e.g., $p_{1a2a2b}(t,t+\tau)$ means that the detector $1a$ clicked at the instant $t$ and the detectors $2a$ and $2b$ clicked simultaneously at the instant $t+\tau$]. Equation (\ref{eq_g2c}) is valid for $\tau>0$ only, since the simultaneous click of $j$ and $j'$ will occur after the single click of $i$. From Eq. (\ref{eq_g2c}), we calculate $\bar{g}_{122}^{(2)}(\tau)$ from a time average similar to the methods of Sec. \ref{sec_setup}. By considering the arrangement of detectors in our setup [Fig. \ref{fig1}(a)], we have four $\bar{g}_{122}^{(2)}$ functions: $\bar{g}_{1a2a2b}^{(2)}$, $\bar{g}_{1b2a2b}^{(2)}$, $\bar{g}_{2a1a1b}^{(2)}$, and $\bar{g}_{2b1a1b}^{(2)}$. Quantum correlations are verified if $g_{c}^{(2)}<1$; ideally, $g_{c}^{(2)}=0$ for single-photon states, and a value smaller than 1/2 indicates suppression of the second-order components into the single-photon regime \cite{Grangier1986,Loudon1983,Laurat2006, Oliveira2014,Felinto2005}.

Figure \ref{fig11} displays the $g_{c}^{(2)}(\tau)$ functions for two different pumping powers. For comparison, we added the second-order correlations functions, $g^{(2)}_{1b1a}$ (lime green) and  $g^{(2)}_{2b2a}$ (gold). The correlation peaks in the cross-correlations functions leading to $R>1$ also indicate regions of suppression of $g_{c}^{(2)}(\tau)$ with respect to the corresponding autocorrelation function. However, we were not able to observe  $g_{c}^{(2)} < 1$ for our experimental data. In many systems, the reduction of optical excitation power makes it possible to achieve the single-photon regime, for example, in the DLCZ protocol \cite{Duan2001}. This procedure here, however, did not change significantly the shape of the $g_{c}^{(2)}(\tau)$ function. The reason for this comes from the small degree of Cauchy-Schwarz violation in our system \cite{Araujo2022}. In a paper employing the DLCZ-protocol scheme, for example, Laurat \textit{et al.} \cite{Laurat2006} observed a violation of the classical condition $g_{c}^{(2)}  \geq 1$ only when the second-order cross-correlation function was on the order or larger than 5, which is almost 1.6 times greater than the correlations obtained in our two-level system \cite{Araujo2022}.

\section{Slow classical regime}
\label{sec_slow}

In the limit where $\tau \rightarrow \infty$, the system tends to become uncorrelated. Therefore, the joint probability $p_{ij}(t,t+\tau)$ can be factored as $p_{ij}(t,t+\tau)=p_i(t)p_j(t+\tau)$. Thus, the correlation function in this limit becomes
\begin{gather}
    \lim_{\tau \rightarrow \infty} g_{ij}(t,t+\tau) =  \lim_{\tau \rightarrow \infty} \frac{p_{ij}(t,t+\tau)}{p_i(t)p_j(t+\tau)} = 1. 
\end{gather}
The timescale to reach this limit characterizes the slow classical regime, which in our system occurs at tens of $\mu$s (see Figs. \ref{fig12} and \ref{fig13}).

\subsection{Single-beam experiment}\label{sec:single_beam}

\begin{figure}[h]
\centering
\includegraphics[width=8.5 cm,height= 5. cm]{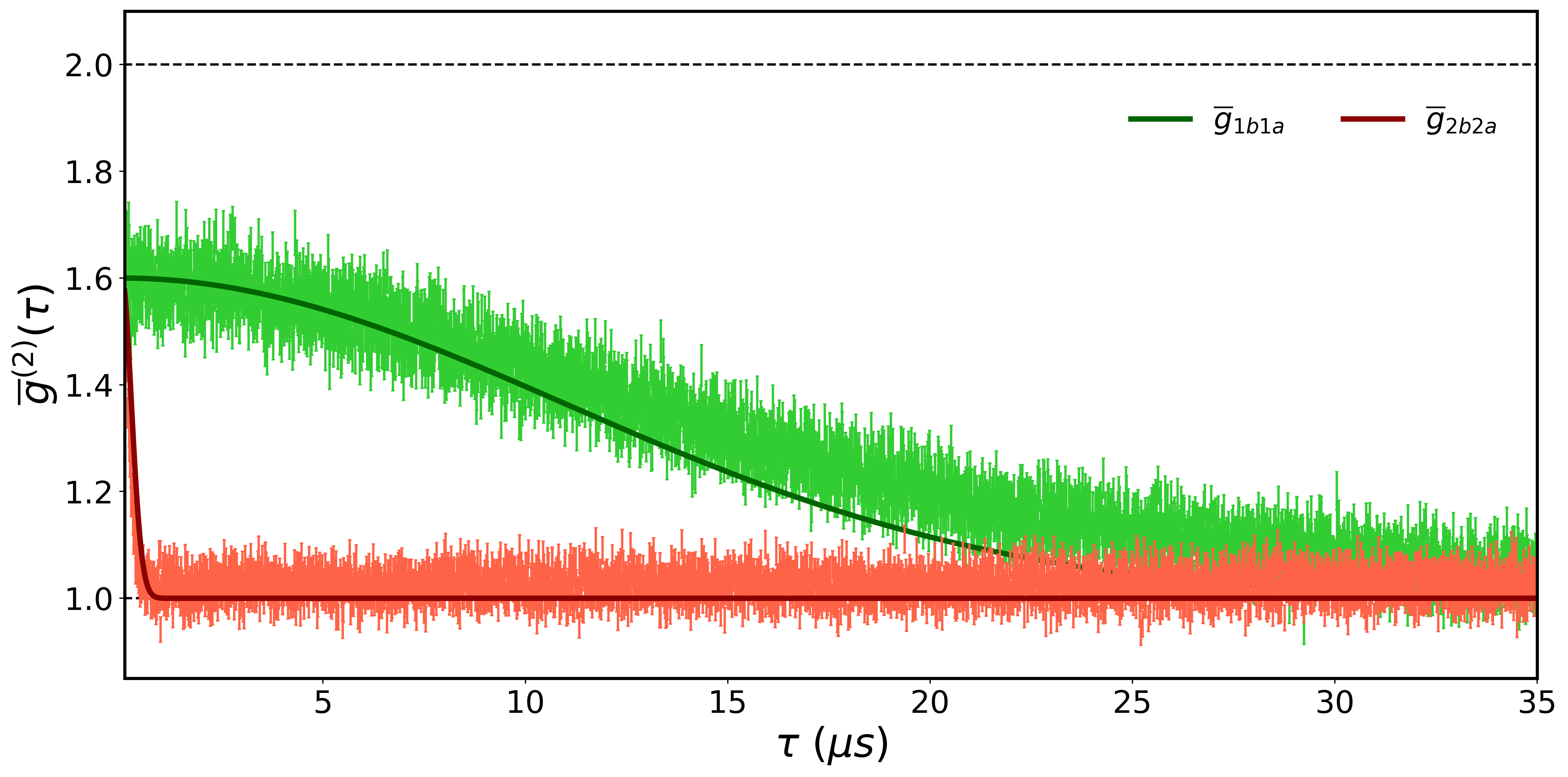}
\caption{Autocorrelation functions $\bar{g}_{1b1a}(\tau)$ (green) and $\bar{g}_{2b2a}(\tau)$ (red) in the configuration with only one excitation beam. The continuous curves represent fittings from Eq. (\ref{gaussian_fit}). The global fit of the data yields to a temperature of $T = 300 \pm 10$ $\mu$K, and a scattering angle $\theta_1 = 2.7 \pm 0.1^{\circ}$ and $\theta_2 = 180^{\circ} -\theta_1$. Measurement was carried out with $P=310$ $\mu$W and $\Delta =+50\Gamma$.}
\label{fig12}
\end{figure}

In order to verify the dependence of the decay time on the long-term scale with the detection angle, we blocked one of the excitation beams (see Fig. \ref{fig3}). Figure \ref{fig12} shows the autocorrelation functions  $g_{11}$ and $g_{22}$ for one of the excitation beams blocked. Also, there are no cross-correlations between fields 1 and 2, so $g_{12}=1$ (uncorrelated) for all $\tau$ values. For better visualization, they are not shown in Fig. \ref{fig12}. This behavior is expected, as we no longer have a phase-matched process with this spatial configuration of a single excitation beam only. The photons detected by APDs $D_{1a}$ and $D_{1b}$ have a scattering angle $\theta=3.0 \pm 0.3^{\circ}$, and for detectors $D_{2a}$ and $D_{2b}$, we have a scattering angle of $180^{\circ}-\theta=177\pm 0.3^{\circ}$. These different scattering angles directly impact the characteristic times for the decay of the autocorrelation functions $\bar{g}_{1b1a}(\tau)$  and $\bar{g}_{2b2a}(\tau)$, as observed in Ref. \cite{Moreira2021}.

The temporal behavior of the autocorrelation functions $\bar{g}_{1b1a}(\tau)$  and $\bar{g}_{2b2a}(\tau)$ can be well described by a Gaussian-like decay \cite{Almeida2016,Moreira2021}

\begin{equation}\label{gaussian_fit}
    g^{(2)}(\tau)=1+A\;e^{-(\tau/\tau_D)^2},
\end{equation}

with the Doppler decay time given by $\tau_{D}=\Lambda/\sqrt{2}\pi u$, where $\Lambda=\lambda/[2\sin(\theta/2)]$ is the spatial period of the coherence grating created between the excitation beam and the emitted mode, with $\lambda$ being the wavelength and $\theta$ the scattering angle, $u=\sqrt{2k_BT/m}$ being the most probable velocity of atoms of mass $m$ at a temperature $T$, and $k_B$ the Boltzmann constant. The parameter $A$ adjusts the value of the autocorrelation functions at $\tau=0$. While theoretically we expect  $A=1$, in order to satisfy the statistics of thermal fields $g^{(2)}(0)=2$, noise and angular mismatch may reduce the value of $A$. After finding the value for $A$, we perform a global fitting of our experimental data, using the temperature $T$ of the atomic cloud and the scattering angle $\theta_1$ (corresponding to the light detected by APDs $D_{1a}$ and $D_{1b}$) as global adjustment parameters [note that $\theta_1$ is the angle $\theta$ included in the parameter $\tau_D$ of Eq. (\ref{gaussian_fit})]. The angle $\theta_2$ (corresponding to the light detected by APDs $D_{2a}$ and $D_{2b}$) is the supplementary of $\theta_1$, i.e., $\theta_2 = 180^{\circ} -\theta_1$. This relation between the angles is fixed by measuring the quantum correlation in the SFWM signal right before blocking one of the beams. The fit yields to a temperature of $T = 300 \pm 10$ $\mu$K, and a scattering angle $\theta_1 = 2.7 \pm 0.1^{\circ}$, $A_1 = 0.60 \pm 0.02$, and $A_2 = 0.62 \pm 0.02$. It is worthwhile to emphasize that we have assumed $\theta_1$ as a fitting parameter, but we restrict its fitting to return values inside the error bar of $0.3^{\circ}$. The broad qualitative agreement we obtained with this procedure indicates that the angular difference between the two detections
leads to the appearance of two, very distinct, slow decays.
%\end{itemize}

Additionally, we noted in Ref.  \cite{Marinho2022thesis} that the autocorrelation functions in Fig. \ref{fig12} decay with nearly the same unnormalized probabilities, at zero delays, for $\theta = 3^{\circ}$ (with decay time $\tau_{D1}$) and for $\theta = 177^{\circ}$ (with decay time $\tau_{D2}$). The slight variations are caused by detectors with slightly different efficiencies. Based on these observations, we will propose in the next subsection an empirical modification in order to globally describe the decay in the correlation functions, both on fast and slow timescales.

\subsection{Full empirical expression}\label{sec:complete_empirical_expression}

\begin{figure}[h]
\centering
\includegraphics[width=8.5 cm,height= 12 cm]{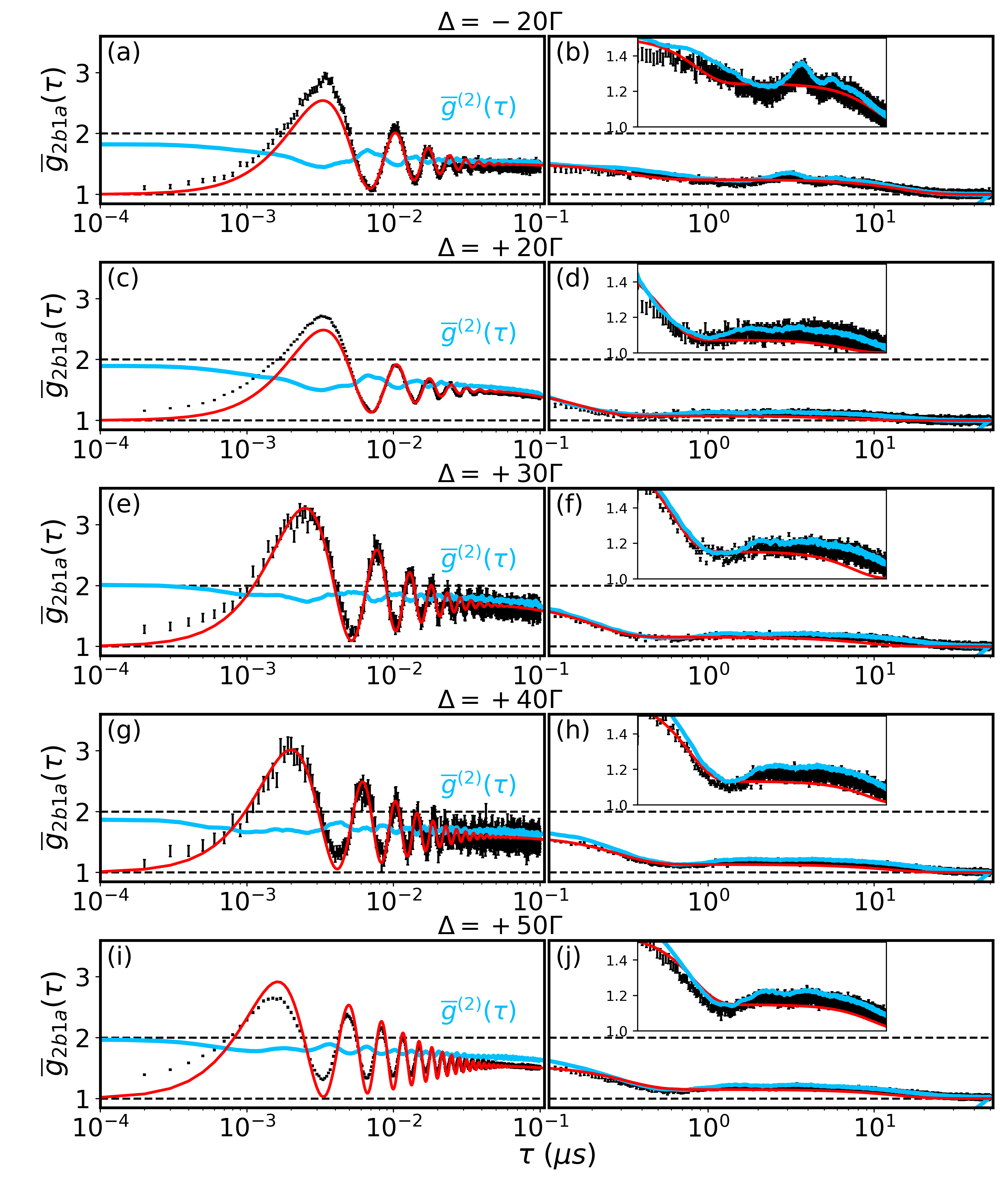}
\caption{Cross-correlation $\bar{g}_{2b1a}(\tau)$ (black) and autocorrelation $\bar{g}^{(2)}(\tau)$ (blue), defined in Appendix \ref{ap2} as the convolution between $\bar{g}_{1b1a}(\tau)$ and $\bar{g}_{2b2a}(\tau)$, including fast and slow time regimes, with the $\tau$ axis in a logarithmic scale. Other combinations of cross-correlation functions give similar results. The black dots are the experimental data, and the red curves represent the theoretical empirical fit from Eq. (\ref{eq_fit_tau_longo}). Measurements were performed with fixed power $P=350$ $\mu$W and varied detunings of $-20$, $+20$, $+30$, $+40$, and $+50\Gamma$. The left (right) panels are data plotted with bin = 1 (bin = 100). These bins correspond to
temporal resolutions of 0.1 and 10 ns, respectively.}
\label{fig13}
\end{figure}

In order to adjust the experimental data with the theory on both fast and slow timescales, and based on the results of Sec. \ref{sec:single_beam}, we empirically write the cross-correlation function as	

\begin{gather}
g_{12}^{emp}(\tau)=1+\big(4f/\pi^2)\Big(1+e^{-\chi\Gamma\tau}-2\cos(\Delta_{fit}\tau)e^{-\chi\Gamma\tau /2}\Big)\nonumber\\  \times \Bigg[ \epsilon e^{-(\tau/\tau_{D1})^2} +(1-\epsilon)e^{-(\tau/\tau_{D2})^2}\Bigg],\label{eq_fit_tau_longo}
\end{gather}
with $f$, $\chi$, $\Delta_{fit}$, $\tau_{D1}$, $\tau_{D2}$, and $\epsilon$ fitting parameters. Our empirical modification considered that a portion $\epsilon$ of the atoms in the ensemble are affected by Doppler broadening with a decay time $\tau_{D1}$ and that a fraction ($1-\epsilon$) of the atoms are affected by Doppler broadening with a decay time $\tau_{D2}$. A similar approach is found in Ref. \cite{Moreira2021}, where the Doppler decay was explained from a first-principles theory in line with Sec. \ref{sec:single_beam}. These different Doppler decay times are associated with scatterings with angles of $\theta = 3^{\circ}$ and $180^{\circ}-\theta =177^{\circ}$, respectively, as mentioned in Sec. \ref{sec:single_beam}. Once we have fixed the angle $\theta_1 = 2.7^{\circ}$, the decay times are determined by the cloud temperature, as in Sec. \ref{sec:single_beam}. The parameters $f$, $\chi$, and $\Delta_{fit}$ were adjusted independently by first performing a fitting on the short timescale (up to 50 ns).

\begin{table}[h]
\caption{\label{tab:table1}
A table with fitting parameters including the fast and slow time scales.}
\begin{ruledtabular}
\begin{tabular}{ccccccc}
$\Delta_{exp}$ ($\Gamma$) & $\Delta_{fit}$ ($\Gamma$) & $T$ ($\mu$K) & $\epsilon$ (\%) & $f$
 &$\chi$\\
\hline
-20 & 23.7 & 350  &  48 & 1.24 & 4.4  \\
+20 & 23.2 & 1300  & 14 & 1.26 & 5.3  \\
+30 & 31.7 & 920  &  22 & 1.74 & 4.7  \\
+40 & 39.8 & 475  &  22 & 1.48 & 4.7  \\
+50 & 49.6 & 395  &  27 & 1.34 & 4.0  \\
\end{tabular}
\end{ruledtabular}
\end{table}

In Fig. \ref{fig13}, we display the correlation function $\bar{g}_{2b1a}(\tau)$  with the $\tau$ axis in a logarithmic scale. All other cross-correlation functions also exhibit similar behavior. The black dots are the experimental data, and the red curves represent the theoretical empirical fits from Eq. (\ref{eq_fit_tau_longo}). The blue curves $g^{(2)}(\tau)$ are defined in Appendix \ref{ap2} as the convolution between $\bar{g}_{1b1a}(\tau)$ and $\bar{g}_{2b2a}(\tau)$ and thus are independent of detector labels. Comparing the cross- and autocorrelations we observe a decoupling between the fast and slow time regimes, with the accelerated decay affecting more notably the cross-correlation in the quantum fast regime, while the decay in the classical slow regime is almost the same for both second-order functions. We expect, then, to be able to build classical or semiclassical
theories for fields with thermal statistics of photons to explain in more detail such slow temporal evolution \cite{Moreira2021, Capella2022}, which we approximate for now by the composite Gaussian decays of Eq. (\ref{eq_fit_tau_longo}). The left panels use bin = 1 and the right panels use bin = 100. Bin = 1 and 100 correspond to temporal resolutions of 0.1 and 10 ns, respectively. This change in resolution was necessary because storing information on a timescale of tens of $\mu$s using a resolution of only 0.1 ns had a prohibitive computational cost. We can observe a good fit of the theoretical-empirical expression to the experimental data. Table \ref{tab:table1} displays the corresponding values of the fitting parameters for the different detunings $\Delta$ considered. The differences between the fitted detunings and the actual parameter are on the order of what was reported before in Ref. \cite{Araujo2022}. As for the parameters $f$ and $\chi$, they are consistent with fluctuations in the quality of alignment and atom numbers in the trap over the long period of time that was taken to build Fig. \ref{fig13}, between $3$ and $4$ months. These are all quantities connected to the short-time dynamics. For the long-time decay, we have a good qualitative agreement, with reasonable timescales consistent with previous results under similar conditions \cite{Moreira2021}. However, the large variation of $\epsilon$ for the negative detuning is just an indication of a broader discrepancy, as we do observe some ringing on the correlation functions for these conditions, see inset for Fig. \ref{fig13} (b). For the other detunings, we did not observe such behavior, but we did observe minima followed by small revivals of the correlation functions in the transition between the two decay timescales, as shown by all insets for the right panels in Fig. \ref{fig13}. These observations were not accounted for by our simplified, empirical approach and may require a first-principles theory of the process to be explained.

\section{Conclusion}
\label{sec_conclusion}

This paper delves deeper into the process of generating nonclassical correlations between photon pairs from an ensemble of pure two-level atoms from spontaneous four-wave mixing. With slight empirical modifications to previous theoretical treatments, we were able to describe the behavior of the correlations both in the fast quantum regime and the thermal regime. In the fast regime, two things deserve more attention, namely, the maximum correlations predicted by the theory and the acceleration in their decay rates. The theories developed in Refs. \cite{Du2007,Wen2007} do not agree with each other concerning the maximum possible values for the cross-correlation functions. One of the reasons for this issue could be that the authors used a completely quantum approach to determine the biphoton generation rate, but they took into account a semi-classical treatment based on rate equations for the Rayleigh scattering. Also, their quantum model does not take into account the possibility of generating collective states, therefore also being unable to explain superradiance-like effects, i.e., the acceleration in decay rates due to collective effects \cite{Dicke1954}. Thus, a purely quantum treatment from first principles for the whole process is still an open question. We believe our present approach will clarify these issues and serve as a guide for future theoretical treatment.

On the other hand, in the slow classical regime, although we have been able to empirically obtain important information from the behavior on this scale, such as the temperature of the atomic cloud, we still need a theory that takes into account the atomic motion and the recoil of atoms after emitting photons to properly explain our results. Such processes are one of the main decoherence mechanisms in our system \cite{Moreira2021} and would explain the decay of correlation functions on this slow scale. Additionally, the contribution from scattering with angles $\theta$ and $180^{\circ}-\theta$ must be considered in the signal in this approach, as both scatterings give rise to different decay times and some unexplained behavior around the transition between the two timescales. 

These correlations in our two-level system without any filtering mechanism have a possible maximum bound \cite{Araujo2022,Wen2007}. This occurs because the biphoton generation rate grows with the square of the number of atoms $N^2$, in the same way as the accidental coincidences (background level), which are proportional to the square of the Rayleigh scattering rate. This happens in such a two-level system because both excitation fields act on the same transition simultaneously, generating noise in both emitted modes. However, we provided already an initial demonstration that spectral filtering may enhance the correlations in our two-level system \cite{Marinho2023}, opening the way for strong cycling transitions to be used in quantum information applications.

\section*{Acknowledgements}
This work was supported by the Brazilian funding agencies Conselho Nacional de Desenvolvimento Cient\'{\i}fico e Tecnol\'{o}gico (CNPq - INCT Grant No. 465469/2014-0), Coordena\c{c}\~{a}o de Aperfei\c{c}oamento de Pessoal de N\'{\i}vel Superior (CAPES - PROEX Grant No.  23038.003069/2022-87), Funda\c{c}\~ao de Amparo \`{a} Ci\^{e}ncia e Tecnologia do Estado de Pernambuco (FACEPE), and Funda\c{c}\~ao de Amparo \`{a} Pesquisa do Estado de S\~{a}o Paulo (FAPESP Grant No. 2021/06535-0). L.S.M. thanks Office of Naval Research (ONR Grant No. N62909-23-1-2014). L.S.M. and M.O.A. contributed equally to this work.

%%%%%%%%%%%%%%%%%%%%%%%%%%%%%%%%%%%%%%%%%%%%
\appendix

\section{Factors that prevent the observation of quantum correlations}
\label{ap1}

\begin{figure}[!ht]
\centering
\includegraphics[width=8.5 cm,height= 6.0 cm]{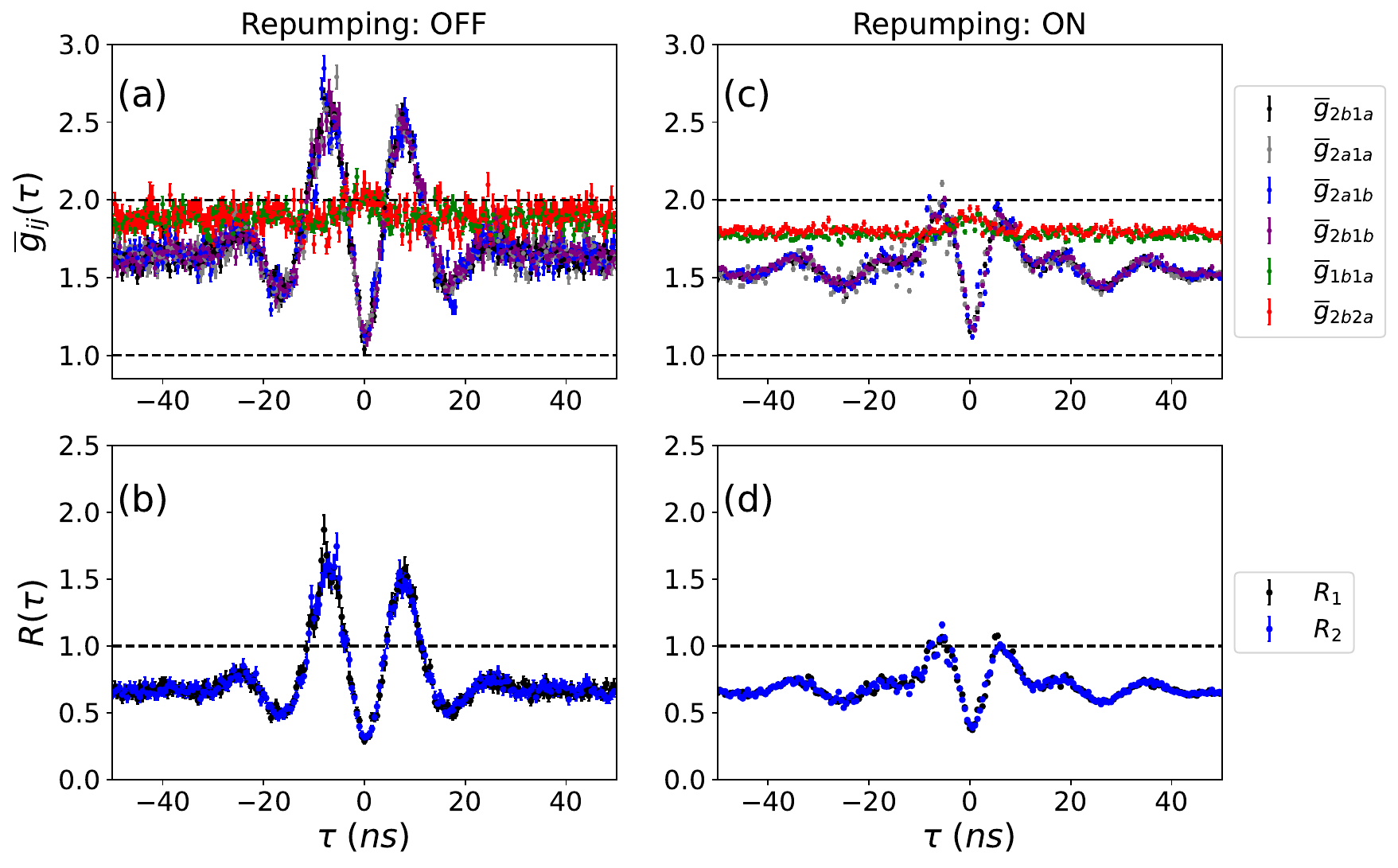}
\caption{Dependence of the correlations $\bar{g}_{ij}$ and the Cauchy inequalities $R_1$  and  $R_2$  with the repumping beams. Panels (a) and (b) [(c) and (d)] denote schemes with the MOT repumping laser beams kept off (on), for $P=19$ $\mu$W and $\Delta=-9\Gamma$. }
\label{figa1}
\end{figure}

\begin{figure}[!ht]
\centering
\includegraphics[width=8.5 cm,height= 6.0 cm]{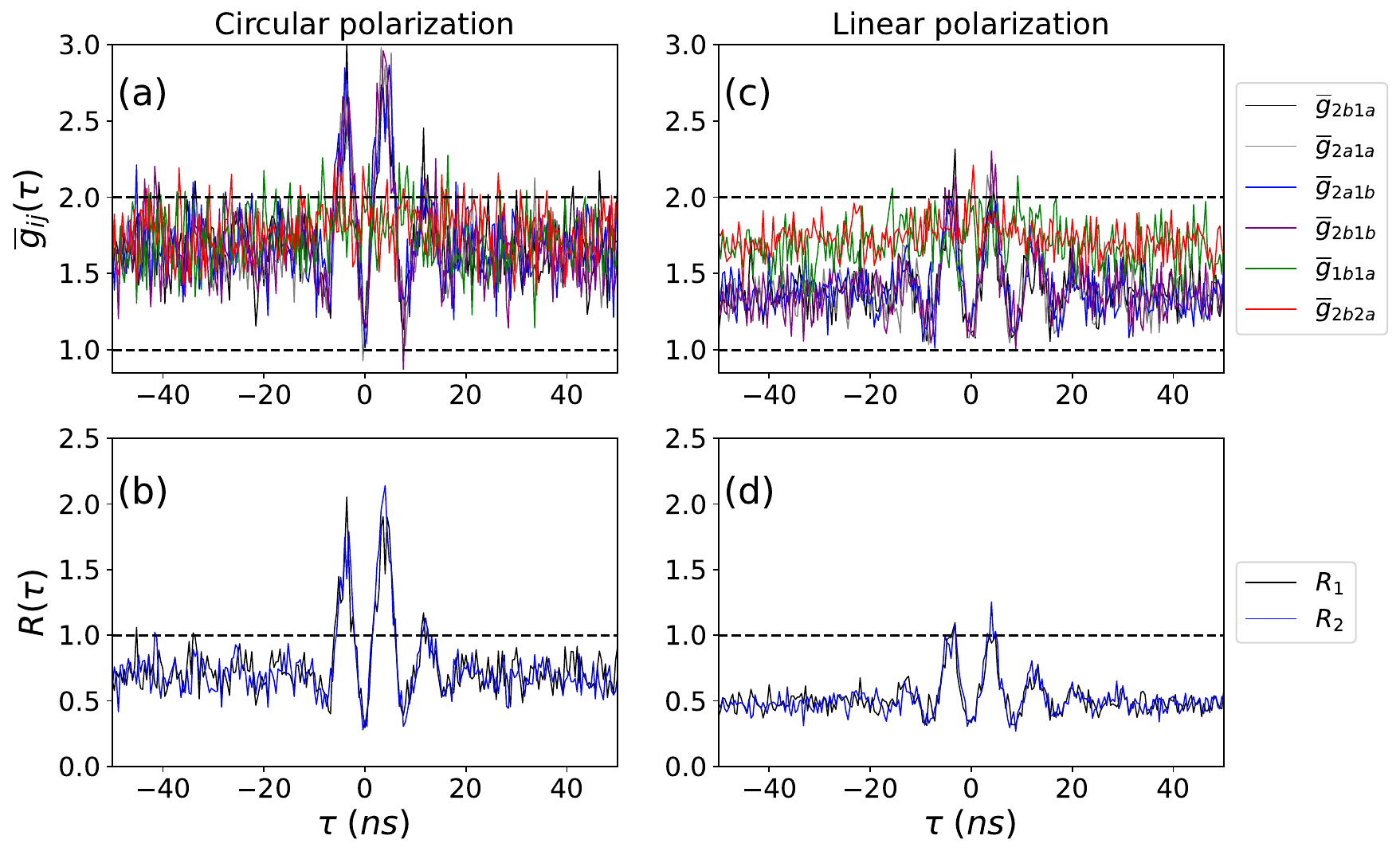}
\caption{Dependence of the correlations $\bar{g}_{ij}$ and the Cauchy inequalities $R_1$  and  $R_2$  with the polarization of the excitation beams. Panels (a) and (b) [(c) and (d)] denote schemes with circular polarization (linear), for $P=+24$ $\mu$W and $\Delta=20\Gamma$. }
\label{figa2}
\end{figure}

As pointed out in the main text and in Ref. \cite{Araujo2022}, the main experimental factor that degrades the cross-correlations is misalignment between the four modes of the SFWM process, and between them and the atomic
cloud. Another critical issue is to decrease as much as possible the crossing of the detection modes with the excitation modes in any other material media, like the windows of the vacuum chamber and other transparent optical elements, as this may lead to spurious scattering of excitation light into the detection modes. These are broad principles that need to be followed in any SFWM experiment exploring quantum correlations between the spontaneously generated photon pairs. There are, however, other factors that are particular for this system of cold atoms. Below we discuss two of these factors that were not fully appreciated by previous groups working in the problem.

To illustrate the impact of the repumping beams on the cross-correlations, we performed a measurement of $\bar{g}_{ij}(\tau)$ and $R$ with and without the repumping beams. In Fig. \ref{figa1}(a) and Fig. \ref{figa1}(b), the absence of the repumping beams (Fig. \ref{fig2}) leads to the observation of quantum correlations with a clear criterion violation. However, when the repumping beams are turned on [Figs. \ref{figa1}(c) and \ref{figa1}(d)], the quantum correlations are suppressed, exactly as observed in Refs. \cite{Wen2007,Du2007}. It is well known that any spurious light or electromagnetic
field kept on during measurements will disturb, to various degrees, the phase between atoms in the collective state, softening signals relying on the state's coherent nature \cite{Felinto2005}. 

We also studied the dependence of the correlations on the polarization of the excitation beams. In Fig. \ref{figa2}(a) and Fig. \ref{figa2}(b), we show the results for the case in which the excitation beams have circular polarization $\sigma^+$. This configuration is responsible for carrying out the optical pumping for the cyclic transition $5S_{1/2}(F=2,m_F=+2) \rightarrow 5P_{3/2}(F^{\prime}=3,m_{F^{\prime}}=+3)$, corresponding to a pure two-level system. On the other hand, in Fig. \ref{figa2}(c) and Fig. \ref{figa2}(d), we show the results for the case in which the excitation beams have horizontal linear polarizations. In this situation, the system can no longer be described as having two pure levels, becoming a mixed state with the atoms distributed among the five Zeeman sub-levels associated with the hyperfine level $F = 2$. The atoms in different Zeeman sub-levels contribute independently through different parametric processes \cite{Felinto2005}, taking the correlations back to the border of the classical region, as observed in Refs. \cite{Du2007,Wen2007}.

\section{Auto-correlation functions behavior}
\label{ap2}

\begin{figure}[htp]
\centering
\includegraphics[width=8.5 cm,height= 12 cm]{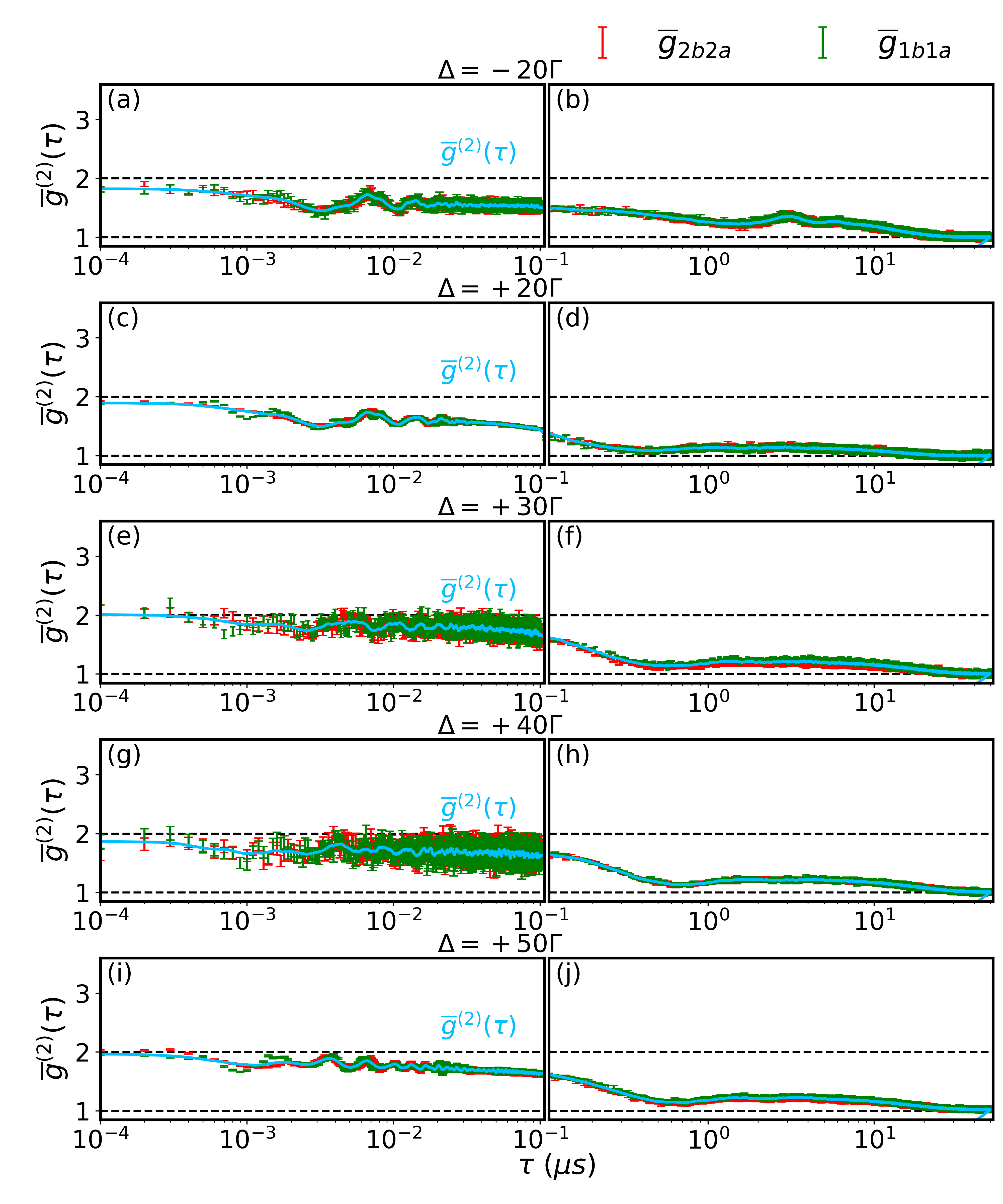}
\caption{Auto-correlation $\bar{g}_{1b1a}(\tau)$ (green) and $\bar{g}_{2b2a}(\tau)$ (red) functions including fast and slow time regimes, with the $\tau$ axis in a logarithmic scale. The blue curves represent the global convolution from the data and are thus independent of detector labels and symbolized by $g^{(2)}(\tau)$. Measurements were performed with fixed power $P=350$ $\mu$W and varied detunings of $-20$, $+20$, $+30$, $+40$, and $+50\Gamma$. The left (right) panels are data plotted with bin = 1 (bin = 100). These bins correspond to
temporal resolutions of 0.1 and 10 ns, respectively.}
\label{figb1}
\end{figure}

In order to demonstrate the behavior of autocorrelation functions in both fast and slow timescales and ensure that the manuscript is self-contained, we display in Fig. \ref{figb1} autocorrelations $\bar{g}_{1b1a}(\tau)$ (green) and $\bar{g}_{2b2a}(\tau)$ (red), with the $\tau$ axis in a logarithmic scale. Since these functions exhibit close similarity, we performed a global convolution to smooth out the data and defined the quantity $g^{(2)}(\tau)$ (blue line) that is independent of the detector labels.

%%%%%%%%%%%%%%%%%%%%%%%%%%%%%%%%%%%%%%%%%%
%\bibliography{references}

\end{document}